\documentclass[preprint,5p,times,twocolumn]{elsarticle}

\usepackage{amssymb}
\usepackage{amsmath}
\usepackage{physics}
\usepackage{graphicx}
\usepackage{bm}
\usepackage{bbm}
\usepackage{hyperref}
\usepackage{todonotes}
\newcommand{\bmk}{\bm{k}}
\newcommand{\bmq}{\bm{q}}
\newcommand{\bmr}{\bm{r}}
\newcommand{\bmrp}{\bm{r}^\prime}
\newcommand{\bmu}{\bm{u}}

\journal{Physics Letters A}

\begin{document}

\begin{frontmatter}

\title{Discrete Time Crystal Phase of Higher Dimensional Integrable Models}

\author[first]{Rahul Chandra}

\author[first]{Analabha Roy\corref{corresponding}}
\ead{aroy@phys.buruniv.ac.in}
\cortext[corresponding]{corresponding author}

\affiliation[first]{addressline={Department of Physics},
    organization={The University of Burdwan},
    city={Bardhaman},
    postcode={713104}, 
    state={West Bengal},
    country={India}}

\begin{abstract}
This paper investigates the possibility of generating Floquet-time crystals in higher dimensions ($d\geq 2$) through the time-periodic driving of integrable free-fermionic models. The realization leads to rigid time-crystal phases that are ideally resistant to thermalization and decoherence.
By utilizing spin-orbit coupling, we are able to realize a robust time-crystal phase that can be detected using novel techniques. Moreover, we discuss the significance of studying the highly persistent subharmonic responses and their implementation in a Kitaev spin liquid, which contributes to our understanding of time translational symmetry breaking and its practical implications.
\end{abstract}

    
\end{frontmatter}

\section{Introduction}
\label{introduction}
Spontaneous Symmetry Breaking (SSB) is a remarkable phenomenon in which a strongly interacting many-body system can synchronize its behavior over long distances and time intervals, leading to macroscopic manifestations. This concept has improved our understanding of various physical systems, including crystalline and magnetic ordering, superfluidity, superconductivity, and the generation of particle masses. The wide range of systems exhibiting this behavior suggests that almost all symmetries have the potential to break. Building on this idea and drawing an analogy to spatial crystals, Wilczek introduced the concept of a "time crystal (TC)"~\citep{PhysRevLett.109.160401}, which refers to a state that spontaneously breaks continuous time-translation symmetry. Further research~\citep{PhysRevA.91.033617,PhysRevLett.116.250401,PhysRevLett.118.030401} has provided a more precise understanding of time translation symmetry breaking (TTSB), including the proof of the non-existence of equilibrium quantum time crystals~\citep{PhysRevLett.111.070402,PhysRevLett.114.251603}. Although TTSB is not possible in equilibrium, it can occur in Floquet systems, which exhibit time-periodic behavior~\citep{PhysRevLett.117.090402}. In such systems, the discrete-time translation symmetry (DTTS) can spontaneously break, resulting in a discrete-time crystal (DTC). This behavior can be observed through broken symmetry, crypto-equilibrium, and rigid long-range order~\citep{10.1063/PT.3.4020}.

As a paradigmatic example of a DTC, consider a spin chain driven by a binary stroboscopic Floquet Hamiltonian that alternates between a spin-echo pulse and a Hamiltonian with interacting spins and a longitudinal field, similar to the Emch-Radin model~\citep{ermodel:emch, ermodel:radin}. In the case where the interactions vanish, the Hamiltonian is trivially integrable. An arbitrarily chosen initial product state undergoes a spin-echo during time evolution, triggering the spins to flip once per Floquet period and then to flip back to the initial state after two periods, resulting in a subharmonic response at half the drive frequency~\citep{PhysRevLett.118.030401}. However, any slight imperfection in the spin-echo pulse immediately disrupts this subharmonic response. Activation of a sufficiently high interaction induces many-body localization (MBL), which preserves the stability of the subharmonic peak, ensuring a stable DTC phase~\citep{PhysRevB.94.085112}.
Thus, the MBL that arises in \emph{non-integrable systems} is a conventional approach to prevent a DTC from melting. In this particular example, MBL effectively contains and restricts the propagation of rotational errors by localizing them. The problem at hand is that all MBL systems are fundamentally prethermal~\citep{huse.mbl,PhysRevResearch.1.033202,PhysRevX.7.011026}. This means that even small instabilities can lead to domain walls that expand over time and finally cause the DTC to reach thermal equilibrium at infinite temperature, as predicted by the Floquet Eigenstate Thermalization Hypothesis~\citep{Mori_2018}.

Multiple approaches are being investigated to protect Time Crystals from thermal decoherence. One such approach is the use of quantum scars, as mentioned in a study by Bull et al. \citep{PhysRevLett.129.140602}. Another approach involves spectral fragmentation through long-range interactions, as discussed in a paper by Pizzi et al.~\citep{Pizzi2021}. In this paper, we propose a protocol that demonstrates the presence of DTC in \emph{integrable systems}. Specifically, we focus on periodically driven systems of "spinless" fermions, which can be created using local fermions in ion traps \citep{Zhang2018} or nonlocal emergent fermions in spin lattices \citep{mbeng_quantum_2020, Chen_2008}. The absence of thermalizing instabilities in integrable systems ensures the persistence of any subharmonics that arise in the dynamics. Our protocol primarily considers bilinear fermionic interactions, specifically Cooper pairs. However, it is important to note that the subharmonic behavior is only observed in a limited number of degrees of freedom within the system. Consequently, new instabilities to perturbations arise, particularly in one-dimensional systems, where the manifold of subharmonics is reduced to a single point. As a result, the rigidity of the DTC is maintained only in higher dimensions, where the manifold is larger. Additionally, detecting individual fermionic degrees of freedom in spin liquids can be challenging, especially if they exhibit nonlocal characteristics. To address this issue, we propose coupling the Cooper pairs to a bosonic field through spin-orbit coupling (SOC). This approach can be implemented in spin liquids by incorporating magnetoelastic coupling into the model, allowing the crystal structure to vibrate \citep{soc:dixon, fiete:phonons}. Consequently, the subharmonics can be observed in momentum-localized phonon degrees of freedom. Various techniques, such as Laser Doppler Vibrometry \citep{doppler:phonons}, atomic force microscopy \citep{Jahng2023}, electron microscopy \citep{Gadre2022}, and pump-probe femtosecond methods \citep{matsuda_fundamentals_2015,Ruello2015-qx}, can be employed to detect these phonons in situ, even when the system is out of equilibrium \citep{ng_excitation_2022}.

This work is organized as follows. In Section~\ref{sec:tcfree}, we obtain a subharmonic response in a generic integrable free-fermion model that is periodically driven. In Section~\ref{sec:stability:detection}, we analyze the rigidity of the ensuing DTC and discuss the difficulties that arise in the realization of the DTC phase. In Section~\ref{sec:soc}, we couple the free-fermion model with a boson field via SOC, producing a \emph{vibrating DTC}, in order to address the realization challenge. In section~\ref{sec:kitaev} , we illustrate the implementation of such a DTC in a Kitaev spin liquid. Finally, we conclude with a discussion and an outlook.

\section{Time Crystal with Free Fermions}
\label{sec:tcfree}
Consider the class of integrable free-fermionic models with Hamiltonian
\begin{equation}
    H  = \sum_{\bmk, -\bmk}\Bigg\{\big(g_0-b_{\bmk}\big)\;\big(c^{\dagger}_{\bmk} c^{\;}_{\bmk}+c^{\dagger}_{-\bmk} c^{\;}_{-\bmk}\big)+\Delta_{\bmk}\;\bigg(c^{\dagger}_{\bmk} c^{\dagger}_{-\bmk}+\text { h.c. }\bigg)\Bigg\}.
    \label{eq:general:hamiltonian}
\end{equation}
Here, $b_{\bmk}=f_{1}\left(J_{i} ; \bmk\right)$ and $\Delta_{\bmk}=f_{2}\left(J_{i} ; \bmk\right)$ are functions of the momenta $\bmk, J_{i}$ denotes Hamiltonian parameters describing specific systems, and $g_{0}$ is a Hamiltonian parameter that we shall vary as a function of time. The sum is carried out over the invariant sectors described by pairs $\pm\bmk$. The fermions are described by the creation (annihilation) operators $c^\dagger_{\bmk}\;\left(c^{\;}_{\bmk}\right)$. Kinetic terms $ c^{\dagger}_{\bmk} c^{\;}_{\bmk}$ correspond to free fermions moving with energy $\left(g_0-b_{\bmk}\right)$, and interacting terms $c^{\dagger}_{\bmk} c^{\dagger}_{-\bmk}$ describe Cooper pairs of strongly correlated fermions with energy $\Delta_{\bmk}$. In the time-independent case, these can easily be mapped to free fermions by a Bogoliubov transformation~\citep{mbeng_quantum_2020}, resulting in $H=\sum_{\bmk}E_{\bmk}\;\gamma^\dagger_{\bmk}\gamma^{\;}_{\bmk}$, with bogolons described by energies $E_{\bmk}=\sqrt{\left(g_0-b_{\bmk}\right)^2 + \Delta^2_{\bmk}}$, and fermionic creation (annihilation) operators $\gamma^\dagger_{\bmk}\;\left(\gamma^{\;}_{\bmk}\right)$ . Such models can be obtained by Jordan-Wigner transformations on spin chains ~\citep{mbeng_quantum_2020}, for example, the TFIM in one dimension and spin liquids, for example, the Kitaev model in a honeycomb lattice of spins in two dimensions~\citep{Chen_2008}. 

\subsection{Subharmonic Response under Periodic Driving}
\label{subsec:subharmonics}
Let us modulate $g_0$ with a square wave so that, during every alternate duty cycle, the Cooper pair interactions vanish and the system evolves as free-fermions with energy $g_1$. Thus, the Hamiltonian $H^{(0)}(t) = \sum_{\bmk, -\bmk} H^{(0)}_{\bmk}(t)$, where
\begin{align}
    H^{(0)}_{\bmk}(t) &= \frac12\;\bigg[1+f(\omega t)\bigg]\;\vqty{H_1}_{\bmk} + \frac12\;\bigg[1-f(\omega t)\bigg]\;\vqty{H_2}_{\bmk} \label{eq:full:hamilt:sc}\\
    \vqty{H_1}_{\bmk} &= \big(g_0-b_{\bmk}\big)\;\big(c^{\dagger}_{\bmk} c^{\;}_{\bmk}+c^{\dagger}_{-\bmk} c^{\;}_{-\bmk}\big)+\Delta_{\bmk}\;\bigg(c^{\dagger}_{\bmk} c^{\dagger}_{-\bmk}+\text { h.c. }\bigg)\nonumber\\
    \vqty{H_2}_{\bmk} &={g_1}\;\big(c^{\dagger}_{\bmk} c^{\;}_{\bmk}+c^{\dagger}_{-\bmk} c^{\;}_{-\bmk}\big),\label{eq:full:hamilt}
\end{align}
and
\begin{equation}
    f(\omega t)=
    \begin{cases}
        +1 & nT \leq t \leq \big(n+\frac12 \big) T \\
        -1 & nT \big(n+\frac12 \big) T < t < \big(n+1\big)T 
    \end{cases}\\
\end{equation}
The function $f(\omega t)$ represents a symmetric unit-amplitude square wave with $50\%$ duty-cycle and frequency $\omega=2\pi/T$.
If we now adjust $\omega$ and focus on the locus of momenta $\bmk_0$ where the following conditions are met,
\begin{equation}
    \label{eq:tc:ks}
    g_0 = b_{\bmk_0}\quad and \quad \omega = \frac{2\Delta_{\bmk_0}}{4m+1},
\end{equation}
with integer $m$, then, for $m=0$, at integer multiples of the time period $T$, the propagator $U(NT) = \displaystyle\prod_{\bmk, -\bmk}U_{\bmk}(NT)$ gets the contributions $U_{\bmk_0}(NT) =  \hat{u}^N_{\bmk_0}$ from the momenta $\bmk_0$, where
\begin{multline}
    \hat{u}^{\;}_{\bmk_0} \equiv \exp{-\frac{ig_1T}{2}\;\bigg(c^{\dagger} _{\bmk_0} c^{\;}_{\bmk_0} + c^{\dagger} _{-\bmk_0} c^{\;}_{-\bmk_0}\bigg)}\\
    \exp{-\frac{i\pi}{2}\;\bigg(c^{\dagger} _{\bmk_0} c^\dagger_{-\bmk_0}+ \rm{h.c.}\bigg)}.
\end{multline}
\begin{figure}[t!]
    \centering
    \includegraphics[scale=3.5]{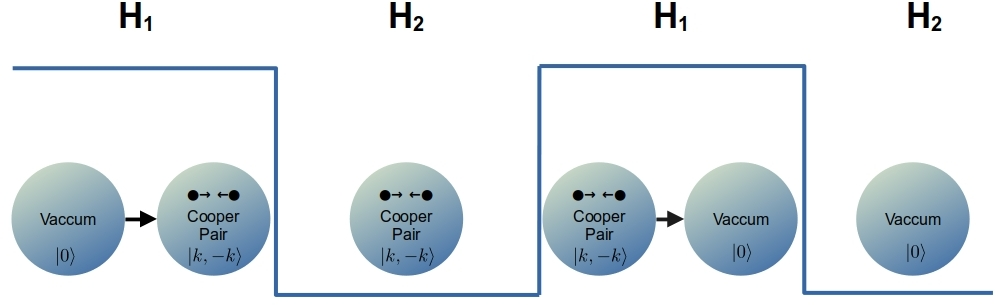}
    \caption{Schematic diagram of the time crystal dynamics at momenta $\bmk_0$ (defined in  equation~\ref{eq:tc:ks}). As the Hamiltonians $H_1, H_2$ (equation~\ref{eq:full:hamilt}) are alternated by the square wave (blue), the initial vacuum produces cooper pairs due to $H_1-$dynamics, which evolve as free-fermions during $H_2-$dynamics, then annihilate back to vacuum during the next cycle. Thus, the dynamics is periodic in two cycles of the square wave.}
    \label{fig:scm:tc}
\end{figure}
Since $\hat{u}^{\;}_{\bmk_0}\ket{0} = -e^{-ig_1T/2} c^\dagger_{\bmk_0}c^\dagger_{-\bmk_0}\ket{0}$, and $\hat{u}^{\;}_{\bmk_0}c^\dagger_{\bmk_0}c^\dagger_{-\bmk_0}\ket{0} = -e^{-ig_1T/2}\ket{0}$, we have $\hat{u}^2_{\bmk_0}\ket{0}\sim\ket{0}$. Thus, if $\ket{\psi(0)}=\ket{0}$, then the probability distributions of the wave functions at momenta $\bmk_0$ repeat themselves every $2T$ times, or at a rate that is half the frequency of the drive \textit{, that is,} with frequency $\Omega = \omega/2$.  The dynamics is shown schematically in figure~\ref{fig:scm:tc}. As the square wave drives the system in time, it evolves under the Hamiltonian $\vqty{H_1}_{\bmk}$ for the first duty cycle, leading to the formation of static Cooper pairs in the $\pm\bmk_0-$ sector, since the kinetic motion is suppressed at those momenta. These Cooper pairs evolve as free-fermions during the second duty cycle under the Hamiltonian $\vqty{H_2}_{\bmk_0}$. During the third duty cycle, evolution under $\vqty{H_1}_{\bmk_0}$ resumes for these Cooper pairs, and they are destroyed to vacuum, which in the $\pm \bmk_0-$ sector only picks up a phase as it evolves under $\vqty{H_2}_{\bmk_0}$ during the fourth duty cycle. Thus, the system repeats every four duty cycles in the $\pm \bmk_0-$ sector or every two periods of the drive\footnote{This dynamics, when translated into the language of spins via the Jordan Wigner Transformation, maps to the spin-echo that was discussed in the introduction.}. If the system is strobed at integer multiples of $T$, the Hamiltonian appears to exhibit time-translational symmetry, but the time-dependent state $\ket{\psi(t)} = \prod_{\bmk}\ket{\psi_{\bmk}(t)}$ of the system has a lower symmetry if its frequency is subharmonic, say $\omega/2$. 

We can also see signatures of this subharmonic in the fermionic off-diagonal correlator in the $\pm \bmk_0-$sector, given by $\mathcal{F}_{\bmk_{0}}^{n}\equiv\expval{c_{\bmk_{0}}^{\dagger} c_{-\bmk_{0}}^{\dagger}+ \mbox{h.c.}}{\psi_{\bmk_{0}}(n T)}$. If we prepare the system in a BCS-like initial state characterized by amplitudes $u^{\;}_{\bmk}, v^{\;}_{\bmk}$, where
\begin{equation}
    \ket{\psi(0)}= \prod_{\bmk, -\bmk}\ket{\psi^{\;}_{\bmk}(0)}, \quad
    \ket{\psi^{\;}_{\bmk}(0)} \equiv \left(u^{\;}_{\bmk} + v^{\;}_{\bmk}c^\dagger_{\bmk}c^\dagger_{-\bmk}\right)\ket{0},	
\end{equation}
then, as long as it is not a Fock state, the contribution of $\pm\bmk_0$ to the wavefunction at $t=T$ 
is given by
\begin{equation}
    \ket{\psi^{\;}_{\bmk_0}(T)} = \hat{u}^{\;}_{\bmk_{0}}\ket{\psi^{\;}_{\bmk_0}(0)} = -i \left(u^{\;}_{\bmk_0} - v^{\;}_{\bmk_0}c^\dagger_{\bmk_0}c^\dagger_{-\bmk_0}\right)\ket{0}.
\end{equation}
From this result it follows that 
\begin{align}
\ket{\psi^{\;}_{\bmk_{0}}(2nT)}&=(-1)^n\ket{\psi^{\;}_{\bmk_{0}}(0)},\nonumber\\
\ket{\psi^{\;}_{\bmk_{0}}((2n+1)T)} &=(-1)^n\ket{\psi^{\;}_{\bmk_{0}}(T)}.
\end{align}
After some simple algebra, these results yield
\begin{equation}
    \mathcal{F}_{\bmk_{0}}^{n}=2 u^{\;}_{\bmk_{0}} v^{\;}_{\bmk_{0}}\;(-1)^{n},
    \label{eq:corrpeaks}
\end{equation}
which leads to a subharmonic peak. 

\subsection{Floquet States}
\label{subsec:floquet:states}
In the analysis of closed isolated quantum systems that undergo periodic drive, the Floquet theory~\citep{Floquet} serves as a foundational framework. Consider a quantum system governed by the time-periodic Schr\"odinger equation $i \pdv{}{t}\ket{\psi(t)} = \hat{H}(t) \ket{\psi(t)}$ with period $T$. Floquet introduced the concept of Floquet states that can be adapted to solve the time-dependent Schrödinger equation. These Floquet states are of the form:
\begin{equation}
    \ket{\Psi_n(t)} = e^{-i\Theta_n t} \ket{\Phi_n(t)},
\end{equation}
where $\Theta_n$ are the quasienergies and the Floquet modes
$\ket{\Phi_n(t)}$ are $T-$periodic functions. Substituting this ansatz into the Schr\"odinger equation yields the Floquet eigenvalue problem,
\begin{equation}
    \hat{H}^{(0)}_F \ket{\Phi^{(0)}_n(t)} = \Theta_n \ket{\Phi^{(0)}_n(t)},
\end{equation}
with the Floquet Hamiltonian $\hat{H}^{(0)}_F = \hat{H}^{(0)}(t) - i \pdv{t}$, which is an operator that is time-averaged over one period of the driving force~\citep{reichl}. Thus, Floquet theory can be used to obtain solutions to the Schr\"odinger equation by diagonalizing the Floquet Hamiltonian. This formalism proves invaluable in a comprehensive understanding of the rich dynamics exhibited by periodically driven quantum systems.

For the dynamics governed by the Hamiltonian in equation~\ref{eq:full:hamilt}, let the $2^M$ Floquet states (for a reciprocal lattice of $2M$ points) be denoted by $\ket{\Phi^{(0)}_n(t)}=\prod_{\bmk, -\bmk}\ket{\phi^\pm_{\bmk}(t)}$, and the quasienergies (in units of $T$) be denoted by $\Theta_n = \sum_{\bmk} \pm\theta_{\bmk}$. Here, $\ket{\phi^\pm_{\bmk(t)}}$ are the two Floquet modes in each $\pm\bmk-$ sector, and the corresponding quasienergies (again in units of $T$) are $\pm\theta_{\bmk}$. For momenta ${\bmk_0}$ in equation~\ref{eq:tc:ks}, the Floquet modes at $t=T$ are simply products of
$\ket{0},c^\dagger_{\bmk} c^\dagger_{-\bmk}\ket{0}$. More generally, the Floquet eigensystem is obtained by solving the following eigenvalue equations.
\begin{equation}
    H^{(0)}_{F,\bmk}\ket{\phi^\pm_{\bmk}(t)} \equiv \Bigg[H^{(0)}_{\bmk}(t) - i \pdv{}{t}\Bigg]\ket{\phi^\pm_{\bmk}(t)} =\pm \frac{\theta_{\bmk}}{T} \ket{\phi^\pm_{\bmk}(t)}.
\label{eq:se}    
\end{equation}
Next, we define the following quantities.
\begin{align}
    A_{\bmk} &= e^{-i g_{1} T / 2}\Bigg\{\cos(\frac{E_{\bmk} T}{ 2})-i n_{3 {\bmk}}\; \sin(\frac{E_{\bmk} T}{ 2})\Bigg\} \;,  \nonumber \\ 
    B_{\bmk} &= -i n^{\;}_{1{\bmk}}\;e^{i g_{1} T / 2}\; \sin(\frac{E_{\bmk} T}{ 2})\;,\nonumber \\
    n^{\;}_{3 {\bmk}} &= \left(g_{0}-b_{\bmk}\right) / E_{\bmk} \;,\quad n^{\;}_{1 {\bmk}} = \Delta_{\bmk} / E_{\bmk} .
\end{align}
The $T$-periodic Floquet modes at $t=nT$ can be expanded to $\ket{\phi^\pm_{\bmk}(nT)} = u_{\bmk}^{\pm}\ket{0} + v_{\bmk}^{\pm}\ket{\bmk, -\bmk}$, where $\ket{\bmk, -\bmk} = c^\dagger_{\bmk}c^\dagger_{-\bmk}\ket{0}$ is a single Cooper pair state that has been created at momenta $\pm \bmk$. The quasienergies and Floquet modes at $t=nT$ can be obtained by solving equation~\ref{eq:se} at $t=T$, yielding
\begin{equation}
    \theta_{\bmk}=\arccos \left[\frac12 \left(A_{\bmk}+A_{\bmk}^{*}\right)\right],
    \label{eq:thetak}
\end{equation}
as well as the amplitudes $u^{\pm}_{\bmk}, v^{\pm}_{\bmk}$,
\begin{align}
        u_{\bmk}^{+}&=B_{\bmk} / D_{\bmk}^{+}=-v_{\bmk}^{-*}(T)  \nonumber \\
        v_{\bmk}^{+}&=\left(\lambda_{\bmk}^{+}-A_{\bmk}\right) / D_{\bmk}^{+}=u_{\bmk}^{-*}(T).
\end{align}
Here, the following quantities are used:
\begin{equation}
\lambda_{\bmk}^{\pm} =e^{\pm i \theta_{\bmk}}\quad and \quad
D_{\bmk}^{\pm} =\sqrt{\abs{\lambda_{\bmk}^{\pm}-A_{\bmk}}^{2}+\abs{B_{\bmk}}^{2}}.
\end{equation}

Floquet modes at all times $t$ can be obtained by solving the differential equation resulting from the substitution of $\theta_{\bmk}$ in equation~\ref{eq:thetak} into~\ref{eq:se}. This yields
\begin{align}
    \ket{\phi^\pm_{\bmk}(t)} &= u_{\bmk}^{\pm}(t)\ket{0} + v_{\bmk}^{\pm}(t)\ket{\bmk, -\bmk},\nonumber\\
    u^\pm_{\bmk}(t) &= \frac12\;u^\pm_0(\bmk,t)\;\bigg[1+f(\omega t)\bigg] + \frac12\;u^\pm_1(\bmk,t)\;\bigg[1-f(\omega t)\bigg],  \nonumber \\
    v^\pm_{\bmk}(t) &= \frac12\;v^\pm_0(\bmk,t)\;\bigg[1+f(\omega t)\bigg] + \frac12\;v^\pm_1(\bmk,t)\;\bigg[1-f(\omega t)\bigg],
    \label{eq:floquet:modes}
\end{align}
and
\begin{align}
    u^\pm_0(\bmk,t) &=\frac12 \Bigg[n^{\;}_{1\bmk}\;v^\pm_{\bmk} \left(1 - e^{2 i E_{\bmk} t}\right) + u^\pm_{\bmk}\;\bigg\{
    e^{2iE_{\bmk}t}\left(1-n^{\;}_{3\bmk}\right)\nonumber\\ 
    &\quad + \left(1+n^{\;}_{3\bmk}\right)
    \bigg\}\Bigg] \; e^{- i \left(E_{\bmk}\mp\frac{\theta_{\bmk}}{T}\right)t} \nonumber \\
    u^\pm_1(\bmk,t) &= \frac12 \Bigg[n^{\;}_{1\bmk}\; v^\pm_{\bmk} \left(1 - e^{ i E_{\bmk} T}\right) + u^\pm_{\bmk}\; \bigg\{
    e^{iE_{\bmk}T}\left(1-n^{\;}_{3\bmk}\right) \nonumber\\
    &\quad + \left(1+n^{\;}_{3\bmk}\right)
    \bigg\}\Bigg]\;e^{-i \left(g_1\pm\frac{\theta_{\bmk}}{T}\right)t}\;e^{-i\left(E_{\bmk}-g_1\right)T/2}\nonumber\\
    v^\pm_0(\bmk,t) &= \Bigg[\frac12 n^{\;}_{1\bmk}\;u^\pm_{\bmk}\left(1-e^{2iE_{\bmk}t}\right)-n^{\;}_{3\bmk}\;v^\pm_{\bmk} \nonumber\\
    &\quad+ \frac{v^\pm_{\bmk}\;n^2_{1\bmk}}{2\left(1-n^{\;}_{3\bmk}\right)}\left(1+e^{2iE_{\bmk}t}\right)\Bigg]\;e^{- i \left(E_{\bmk}\mp\frac{\theta_{\bmk}}{T}\right)t} \nonumber\\
    v^\pm_1(\bmk,t) &=\Bigg[\frac12 n^{\;}_{1\bmk}\;u^\pm_{\bmk}\left(1-e^{iE_{\bmk}T}\right)-n^{\;}_{3\bmk}\;v^\pm_{\bmk} \nonumber\\
    &\quad + \frac{v^\pm_{\bmk}\;n^2_{1\bmk}}{2\left(1-n^{\;}_{3\bmk}\right)}\left(1+e^{iE_{\bmk}T}\right)\Bigg]\;e^{-i \left(g_1\pm\frac{\theta_{\bmk}}{T}\right)t}\;e^{-i\left(E_{\bmk}+g_1\right)T/2}.
    \label{eq:floquet:amplitudes}
\end{align}
These amplitudes can be used yield the complete solution to the time-periodic Hamiltonian in equation~\ref{eq:full:hamilt:sc} for all initial conditions and for all times.

\section{Rigidity and Realization of the Time Crystal Phase}
\label{sec:stability:detection}
Based on the findings of the previous section, it is evident that if we initiate the system from a non-Floquet state at $t=0$, then at the $\pm\bmk_0-$ sector, the driven system repeats at a frequency that is half of the drive frequency. Consequently, if we sample the system at integer multiples of the time period, it will appear to be time-invariant, although the states at $\pm\bmk_0$ will still vary with time. This implies that there is a discrete violation of time translational symmetry. However, for this violation to be considered as a time crystal phase of matter, it must be stable against perturbations, or in other words, it must be rigid and realistically detectable.

The issue that arises with rigidity is that the momentum values $\bmk_0$ are highly dependent on the external drive parameters (in this case, represented by $g_0, \omega$). The stability of the condition in equation~\ref{eq:tc:ks} will vary depending on the dimensionality of the system under consideration. When the dimensionality is $d=1$, the condition is satisfied for a single momentum within the Brillouin zone. Consequently, any deviation will eliminate this subharmonic and disrupt the rigidity of the crystal. However, when $d>1$, this condition is generally met by all $\bmk_{0}$ that lie on a surface with dimensions $d-1$ within the Brillouin zone of dimensions $d$. The subharmonic dynamics of the correlator at $\bmk_0$ shown in equation~\ref{eq:corrpeaks} remains stable even when there is a change in the Hamiltonian parameters within a certain range where $\Delta_{\bmk_{0}}=\omega/2$ and equation~\ref{eq:tc:ks} is satisfied for at least one $\bmk_{0}$. This guarantees the rigidity of the time-crystal phase, which is characterized by the subharmonic peak. To provide a specific example, let us consider a system with dimension $2$, 
\begin{figure}[t!]
    \includegraphics[width=0.48\textwidth, keepaspectratio]{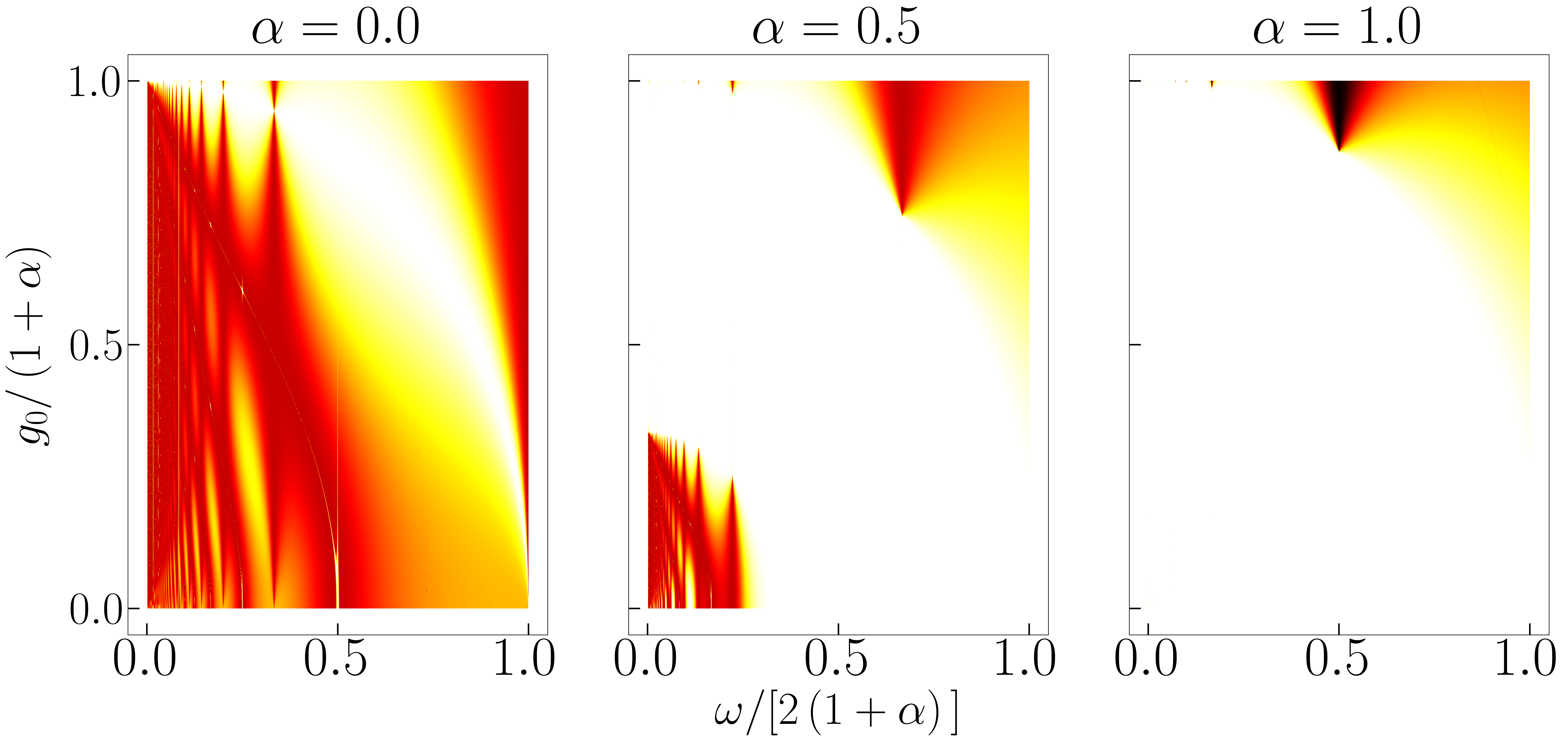}
    \caption{Temporal fluctuations in the $(g_0,\omega)$ plane of the fidelity of a state starting from vacuum $\ket{0}$ and
    sampled at $10000$ integer multiples of $2T=\pi/\omega$, evolving in a 2D DTC with a Kitaev dispersion . The fidelity is evaluated at the momentum $(k_x, k_y)$ where the cost function in equation~\ref{eq:cost:fn} is minimum. The chosen values of anisotropy $\alpha$ are indicated in the panel titles. The left panel, where $\alpha=0$, replicates the 1D case. Fluctuations are zero (white) when the DTC phase is stable}
\label{fig:fluctuations}.
\end{figure}
where $b_{\bmk} = \cos{k_x} + \alpha \cos{k_y}$, $\Delta_{\bmk} = \sin{k_x} + \alpha \sin{k_y}$, with $\alpha \lesssim 1$. These types of dispersions can be easily obtained, for example, in Kitaev spin liquids on a brick lattice ~\citep{Chen_2008}. The analytical results from the previous section can be readily applied to this system, with an additional advantage: The subharmonic state now occurs at the locus of momenta given by the following equations.
\begin{equation}
    g_0 = \cos{k_x} + \alpha\cos{k_y}, \quad \omega = 2\left(\sin{k_x} + \alpha\sin{k_y}\right).
    \label{eq:tcpoint:kitaev}
\end{equation}
Here, minute perturbations in $g_0$ only bring the subharmonic state to a different value of $\bm{k_0}$ in a continuous manifold, thus ensuring the persistence of the subharmonic state. 

The rigidity of the DTC in two dimensions can also be illustrated by numerically evaluating the propagator $U^{\;}_{\bmk}(NT)$ (defined in Section~\ref{sec:tcfree}) at even $N$ and then using it to determine the fidelity of the wave function at integer multiples of $2T$. The fidelity, denoted as $f_{\bmk}(2nT) \equiv \abs{\ip{\psi_{\bmk}(0)}{\psi_{\bmk}(2nT)}}^2$, is a measure of how well the wave function retains its form over time. If there exists a momentum $\bmk_0$ that satisfies equation~\ref{eq:tcpoint:kitaev}, then $f_{\bmk_0}(2nT)=1\;\forall n$. However, if this condition is not satisfied, then $f_{\bmk}(2nT)$ will vary. Therefore, the temporal fluctuations in $f_{\bmk}$ at the momentum closest to satisfying ~\ref{eq:tcpoint:kitaev} serve as a clear indication of the breakdown of the time crystal phase in the $(g_0,\omega)$ plane. For each pair $g_0,\omega$, we employ the trust region method~\footnote{Trust region methods are optimization techniques commonly utilized in machine learning. For more details, see, for example,~\citep{Conn2000}.} from the SciPy library~\citep{2020SciPy-NMeth} to numerically explore the FBZ and identify the momentum $(k_x, k_y)$ that optimally solves the equations~\ref{eq:tcpoint:kitaev} \textit{, that is,} the momentum that minimizes the cost function $\epsilon(k_x, k_y)$, where 
\begin{equation} 
\epsilon(k_x, k_y) \equiv \left(g_0-\cos{k_x} - \alpha\cos{k_y}\right)^2 + \left(\frac{\omega}{2}-\sin{k_x} - \alpha\sin{k_y}\right)^2. 
\label{eq:cost:fn} 
\end{equation} 
If the pair allows for an exact solution to equation~\ref{eq:tcpoint:kitaev}, then the cost function reduces to $0$ only when $(k_x, k_y)=\bmk_0$, resulting in minimal temporal fluctuations in the fidelity at $(k_x, k_y)$. Conversely, when this condition is not met, the cost function minimizes to a nonzero value, and equation~\ref{eq:tcpoint:kitaev} is never exactly satisfied, leading to nonzero temporal fluctuations in fidelity. These fluctuations are depicted in figure~\ref{fig:fluctuations}. The leftmost panel mirrors the one-dimensional scenario, as $\alpha=0$. The results indicate that the time crystal phase is relatively unstable in $1D$, where small deviations in $g_0$ or $\omega$ can transform it from stable to unstable state. However, for two-dimensional cases, the time crystal phase remains stable over a wide range of contiguous $g_0, \omega$ values. Consequently, the phase exhibits rigidity throughout most of the $(g_0, \omega)$ plane. Further elaboration on this rigid phase is provided in Subsection~\ref{subsec:phasediag}, where the phase diagram is extensively discussed.

The next issue is that of realization. Physical quantities that can be measured directly in spin-liquids, such as spin correlations, involve integrals over all fermionic momentum sectors in the Brillouin zone. For example, obtaining the spatial $4-$fermion correlation for the $2-$dimensional problem can be done via the Jordan-Wigner transformation and the application of Parseval's Theorem. This yields, in the continuum limit ~\citep{mbeng_quantum_2020}
\begin{multline}
\expval{\sum_{\bmr,\bmrp}c^\dagger_{\bmr}c^{\;}_{\bmr}c^\dagger_{\bmrp}c^{\;}_{\bmrp}}_{t=nT} = \frac{1}{4\pi^2}\int_{FBZ} \mathrm{d}\bmk\; \Bigg[1 + 4\; \abs{u^+_{\bmk}(T)}^2\;\abs{v^+_{\bmk}(T)}^2\\
\bigg\{-1+\cos(2n\theta_{\bmk})\bigg\} \Bigg].
\label{eq:zcorr}
\end{multline}
Here, the real-space fermion operators $c^\dagger_{\bmr}, c_{\bmr}$ are defined such that 
\begin{equation}
    c^\dagger_{\bmk} \equiv \frac{1}{\sqrt{N}}\sum_{\bmr}c^\dagger_{\bmr}\;e^{-i\bmk\cdot\bmr}
    \quad and \quad 
    c^{\;}_{\bmk} \equiv \frac{1}{\sqrt{N}}\sum_{\bmr}c_{\bm{r}}\;e^{i\bmk\cdot\bmr}.
    \label{eq:crcrd}
\end{equation}
However, the integral in equation~\ref{eq:zcorr} spans the entire Brillouin zone of the system ~\citep{Chen_2008}, and it is expected that the contribution of the subharmonic signal at the single momentum $\bmk_0$ disappears at the thermodynamic limit. Even if that is not the case, the only way a subharmonic would manifest in equation~\ref{eq:zcorr} is through the $n-$dependent cosine term on the RHS. This will not survive long, since that contribution to the integral oscillates so rapidly when $n\rightarrow\infty$ that the integral averages out, leaving behind a constant term without a subharmonic signature. Thus, the subharmonic response does not contribute to measurable correlations in any meaningful way. 

In the following section, our aim is to address this realization issue by enhancing the influence of $\bmk_0$ on the dynamics. To achieve this, we propose perturbative coupling of the system to a bosonic particle bath while ensuring the preservation of unitarity in the dynamics. By allowing all Cooper pairs to interact with the bosons, the subharmonic in the $\pm\bmk_0-$sector can be detected through the resulting dynamics of the phonons, rather than relying on observables in the unperturbed system.

\section{The Vibrating Time Crystal: Spin-Orbit coupling}
\label{sec:soc}
To solve the realization problem presented in the previous section, we analyze the dynamics of quantized deformations of the underlying lattice from its equilibrium configuration. This means that a Cooper pair in the undeformed system can scatter off the orbital degrees of freedom of these lattice vibrations, producing subharmonic responses in the dynamics of the resultant excitations that are similar to those in the undeformed system.

\subsection{The Spin-Orbit Coupled Hamiltonian}
Suppose that we have an integrable system where the "spin" degrees of freedom $\bmk$ are coupled to orbital degrees of freedom with momentum $\bm{q}$. This kind of system can be obtained if the underlying crystal structure in the spin liquid vibrates and regular spin-orbit coupling (SOC) is factored into the model. For simplicity, we focus on the deformations introduced by magnetoelastic coupling and assume that the SOC is perturbatively coupled ~\citep{soc:dixon,fiete:phonons,soc:phonons}. 

The unperturbed system is given by $H^{(0)}(t)$ in equation~\ref{eq:full:hamilt:sc}. If the lattice deforms from its equilibrium configuration, the Hamiltonian changes to $H(t) = H^{(0)}(t) + V$. Let us rewrite in coordinate space the $\vqty{H_1}_{\bmk}$ term in~\ref{eq:full:hamilt:sc}, which manifests during the first half-period of $H(t)$, using the creation and annihilation operators defined in equation~\ref{eq:crcrd}. This yields
\begin{align}
    H^{(0)}(t) &= \frac{1}{2} \bigg[1+f(\omega t)\bigg]\Bigg\{2g_{0} \sum_{\bmr}c^\dagger_{\bmr}c_{\bmr} - \sum_{\bmr_1\bmr_2}\; 2b\left(\bmr_1-\bmr_2\right) c^\dagger_{\bmr_1}c^{\;}_{\bmr_2}  \nonumber \\
    &+ \sum_{\bmr_1\bmr_2} \frac{1}{2} \bigg[\Delta\left(\bmr_1-\bmr_2\right)\; c^\dagger_{\bmr_1}c^\dagger_{\bmr_2} + \rm{h.c}\bigg] \Bigg\}  \nonumber \\
    &+ \frac{g_1}{2} \bigg[1-f(\omega t)\bigg]\; \sum_{\bmk, -\bmk}\left(c^\dagger_{\bmk} c^{\;}_{\bmk} + c^\dagger_{-\bmk} c^{\;}_{-\bmk}\right).
    \label{eq:unpert:hamilt}
\end{align}
Here,
\begin{equation}
    b(\bmr) \equiv \frac{1}{N}\sum_{\bmk,-\bmk}b_{\bmk}\;\cos{\big(\bmk\cdot\bmr\big)}  \quad and \quad
    \Delta(\bmr) \equiv \frac{1}{N}\sum_{\bmk,-\bmk}\Delta_{\bmk}\;e^{-i\bmk\cdot\bmr}
    \label{eq:b:delta}
\end{equation}
Now, proceeding in a manner similar to ~\citep{soc:dixon, fiete:phonons}, we make the \textit{nearest-neighbor harmonic approximation}. Let the equilibrium lattice positions be very slightly deformed from $\bmr\rightarrow \bmr+\bmu(\bmr,t)$, where $\norm{\bmu(\bmr, t)}\ll \norm{\bmr}$. In that limit, we can ignore the role of multiphonon processes and approximate the deformations in $b,\Delta$ by Taylor's expansion to the first order.  Next, we make the approximation where the gradients are proportional to the lattice basis vectors $\bmr_\alpha$ and are only significant between two lattice positions that are separated by a basis vector. This yields
\begin{align}
    b\left(\bmr_1-\bmr_2\right) &\rightarrow b\left(\bmr_1-\bmr_2\right) -\kappa \bmr_\alpha\cdot\left(\bmu_1 - \bmu_2\right)\;\delta_{\bmr_1,\bmr_2+\bmr_\alpha}\nonumber\\
    \Delta\left(\bmr_1-\bmr_2\right) &\rightarrow \Delta\left(\bmr_1-\bmr_2\right) + -\lambda \bmr_\alpha\cdot\left(\bmu_1 - \bmu_2\right)\;\delta_{\bmr_1,\bmr_2+\bmr_\alpha}
    \label{eq:deformations}
\end{align}
Here, the stiffness constants $\kappa,\lambda$ are assumed to be very small compared to $N^{-1}\norm{H^{(0)}(t)}$. For simplicity, it is also assumed that they are independent of the basis vectors. We now quantize the orbits $\bmu$ as
\begin{equation}
    \bmu(\bmr) = \frac{1}{\sqrt{N}}\sum_{\bmq,\mu}\frac{\bm{P}_{\bmq,\mu}}{\sqrt{2m\nu_{\bmq,\mu}}}\left(b_{\bmq,\mu} + b^\dagger_{-\bmq,\mu}\right)e^{i\bmq\cdot\bmr}.
    \label{eq:phonon:quant}
\end{equation}
Here, $\bmq,\mu$ represents the momentum and polarization of a boson field (in a physical lattice, these bosons would correspond to quanta of sound, \textit{i.e.}, phonons), $b_{\bmq, \mu}, (b^\dagger_{\bmq, \mu})$ are annihilation (creation) operators of the boson Fock states , $\bm{P}_{\bmq,\mu}$ is the corresponding polarization direction, and $\nu_{\bmq,\mu}$ is the energy-momentum dispersion for the bosons~\footnote{See, for instance,~\citep{marder:ch13}}. Substituting equation~\ref{eq:phonon:quant} into Eqs.~\ref{eq:deformations}, then substituting the result into Eqs.~\ref{eq:b:delta} allows the transformation of equation~\ref{eq:unpert:hamilt}  $H^{(0)}(t)\rightarrow H^{(0)}(t) + V$, where
\begin{equation}
    V= \sum_{\bmk, -\bmk}V_{\bmk} + \sum_{\bmq,\mu}\nu_{\bmq, \mu}\left( \;b^\dagger_{\bmq, \mu} b^{\;}_{\bmq, \mu} +  b^\dagger_{-\bmq, \mu}b^{\;}_{-\bmq, \mu}\right),
\end{equation}
and the scattering potential for each Cooper-Pair, 
\begin{multline}
    V_{\bmk} = \sum_{\bmr_\alpha, \bm{q^\prime},\mu^\prime}\Bigg[\frac{\bmr_\alpha\cdot\bm{P}_{\bm{q^\prime},\mu^\prime}}{\sqrt{2m\nu_{\bm{q^\prime},\mu^\prime}}}\left(b_{\bm{q^\prime},\mu^\prime} + b^\dagger_{-\bm{q^\prime},\mu^\prime}\right)   \Big\{ \kappa\;c^\dagger_{\bm{q^\prime}+\bmk}c^{\;}_{\bmk}\;e^{i\left(\bm{q^\prime}+\bmk\right)\cdot\bmr_\alpha}  \\
    + \frac{\lambda}{2}\left( c^\dagger_{\bmk}c^\dagger_{\bm{q^\prime}-\bmk}\;e^{i\left(\bm{q^\prime}-\bmk\right)\cdot\bmr_\alpha}+\mathrm{h.c}\right) \Big\} \times \left(e^{-i\bm{q^\prime}\cdot\bmr_\alpha}-1\right) + \mathrm{h.c}\Bigg].
    \label{eq:perturbation}
\end{multline}
Equation~\ref{eq:perturbation} can be simplified by absorbing the lattice-dependent terms into constants. First, define the Structure Factor,
\begin{equation}
    S_\alpha(\bmq^\prime,\mu^\prime)\equiv\frac{\bmr_\alpha\cdot \bm{P}_{\bmq^\prime,\mu^\prime}}{\sqrt{2m\nu_{\bmq^\prime,\mu^\prime}}}\left(1-e^{i\bmq^\prime\cdot \bmr_\alpha}\right),
    \label{eq:sfactor}
\end{equation}
as well as basis vector sums
\begin{align}
    \kappa^{\;}_{\bmk,\bmq^\prime,\mu^\prime} &\equiv\kappa\sum_{\bmr_\alpha}e^{i\bmk\cdot \bmr_\alpha}\;S_\alpha(\bmq^\prime,\mu^\prime),\nonumber\\
    \lambda^{\;}_{\bmk,\bmq^\prime,\mu^\prime} &\equiv \lambda\sum_{\bmr_\alpha}e^{-i\bmk\cdot \bmr_\alpha}\;S_\alpha(\bmq^\prime,\mu^\prime).
\end{align}
Next, make the simplifying assumption $\lambda^{\;}_{\bmk,\bmq^\prime,\mu^\prime}\approx \lambda_0 \left(\delta_{\bmq, \bmq^\prime} + \delta_{-\bmq, \bmq^\prime}\right)\;\delta_{\mu\mu^\prime}$. Physically, this means that the deformations in the gap term $\Delta(\bmr)$ resonate only for a particular $\pm\bmq, \mu$. For a lattice with a basis, this could correspond to transverse optical phonon modes in the long-wavelength limit ~\citep{marder:ch13}. Note that both $\bmq$ and $-\bmq$ must contribute to deformations in order to maintain time-reversal symmetry. Furthermore, we note that the $\kappa-$ term will always kill the vacuum state, and the $\ket{\bmk, -\bmk}$ Cooper pair states are its eigenstates with unit eigenvalue $\forall \bmq$. Thus, it will not contribute any $\bmq-$dependent terms to the dynamics, and it can be safely ignored. In that case, we can simplify equation~\ref{eq:perturbation} to yield
\begin{gather}
    V_{\bmk}(\bmq) = \frac12\left(W_{\bmq,\mu}\otimes F_{\bmk,\bmq,\mu} + \rm{h.c.}\right)\label{eq:perturbation:simplified}, \;     W_{\bmq,\mu} =b_{\bmq,\mu} + b^\dagger_{-\bmq,\mu},\nonumber\\
    F_{\bmk,\bmq,\mu} = \frac{\lambda_0}{2}\left( c^\dagger_{\bmk}c^\dagger_{\bmq-\bmk}+c^{\;}_{\bmq-\bmk}c^{\;}_{\bmk} + c^\dagger_{\bmk}c^\dagger_{-\bmq-\bmk}+ c^{\;}_{-\bmq-\bmk}c^{\;}_{\bmk}\right).
    \label{eq:perturbation:gathered}
\end{gather}
We can further simplify the problem by tracing out the bosonic degrees of freedom~\citep{PhysRevB.102.235154}, this technique is equivalent to "integrating out the particle bath" in open quantum systems~\citep{10.1093/acprof:oso/9780199213900.001.0001}. First, we define the operators
\begin{align}
    Q_{\pm \bmq,\bmk} &\equiv \frac{\lambda_{0}}{2\sqrt{\nu_{\bmq}}} c^\dagger_{\bmk} c^\dagger_{\pm \bmq -\bmk} + \frac{\sqrt{\nu_{\bmq}}}{2}b^\dagger_{\bmq} \nonumber \\
    \overline{Q}_{\pm \bmq,\bmk} &\equiv \frac{\lambda^\ast_{0}}{2\sqrt{\nu_{\bmq}}} c^\dagger_{\bmk} c^\dagger_{\pm \bmq -\bmk} + \frac{\sqrt{\nu_{\bmq}}}{2}b_{\bmq}  \nonumber\\
    R_{\pm \bmq,\bmk} &\equiv \frac{\lambda_{0}}{2\sqrt{\nu_{\bmq}}} c^\dagger_{\bmk} c^\dagger_{\pm \bmq -\bmk} + \frac{\sqrt{\nu_{\bmq}}}{2}b_{-\bmq}  \nonumber\\
    \overline{R}_{\pm \bmq,\bmk} &\equiv \frac{\lambda^\ast_{0}}{2\sqrt{\nu_{\bmq}}} c^\dagger_{\bmk} c^\dagger_{\pm \bmq -\bmk} + \frac{\sqrt{\nu_{\bmq}}}{2}b^\dagger_{-\bmq}.
\end{align}
This allows for rewriting the scattering potential in equation~\ref{eq:perturbation:gathered} as
\begin{multline}
    V_{\bmk}(\bmq)= - \frac{|\lambda_0|^2}{2\nu_{\bmq}} c^{\;}_{\bmq-\bmk} c^{\;}_{\bmk} c^\dagger_{\bmk} c^\dagger_{\bmq-\bmk} -  \frac{|\lambda_0|^2}{2\nu_{\bmq}} c^{\;}_{-\bmq-\bmk} c^{\;}_{\bmk} c^\dagger_{\bmk} c^\dagger_{-\bmq-\bmk}  \\
    + \left( Q^\dagger_{\bmq,\bmk}Q_{\bmq,\bmk} + \overline{Q}^\dagger_{\bmq,\bmk}\overline{Q}_{\bmq, \bmk}+ Q^\dagger_{-\bmq,\bmk}Q_{-\bmq,\bmk}+ \overline{Q}^\dagger_{-\bmq,\bmk}\overline{Q}_{-\bmq,\bmk}\right)  \\ 
    + \left( R^\dagger_{\bmq, \bmk}R_{\bmq, \bmk} + \overline{R}^\dagger_{\bmq, \bmk}\overline{R}_{\bmq, \bmk}+ R^\dagger_{-\bmq, \bmk}R_{-\bmq, \bmk}+ \overline{R}^\dagger_{-\bmq, \bmk}\overline{R}_{-\bmq, \bmk}\right).
\end{multline}
Finally, we define a reduced Hamiltonian $H^R(\bmq, t)\equiv  \displaystyle\sum_{\bmk,-\bmk} H^{(0)}_{\bmk}(t) + V^R_{\bmk}(\bmq)=\displaystyle\sum_{\bmk} H^{(0)}_{\bmk}(t) + \Tr_{D\overline{D}}[V_{\bmk}(\bmq)]$, where we perform a partial trace over the Hilbert space spanned by the eigenvalues of the number operators of the composite particles annihilated by $D\overline{D}=Q\overline{Q},R\overline{R}$. This yields a system with all phonon degrees of freedom traced but with an added $4-$fermion interaction renormalized by the phonon energies. The reduced interaction is given by 
\begin{equation}
    V^R_{\bmk}(\bmq) = - \frac{|\lambda_0|^2}{\nu_{\bmq}}\left(c^{\;}_{\bmq-\bmk} c^{\;}_{\bmk} c^\dagger_{\bmk} c^\dagger_{\bmq-\bmk} + c^{\;}_{-\bmq-\bmk} c^{\;}_{\bmk} c^\dagger_{\bmk} c^\dagger_{-\bmq-\bmk}\right).
    \label{eq:reduced:interaction}
\end{equation}

\subsection{Floquet-Bloch Scattering Amplitudes}
\label{subsec:floquet:scattering}
The Scr\"odinger dynamics of the Hamiltonian obtained from Eqs.~\ref{eq:unpert:hamilt} and~\ref{eq:reduced:interaction} can be solved by treating it as a problem of scattering through a time-periodic potential. Here, $V$ is the time-independent contribution of particle scattering, and $H^{(0)}(t)$ is the noninteracting part, which is also time-periodic. The noninteracting dynamics can be completely described by the quasi-stationary Floquet states that have been discussed in Section~\ref{subsec:floquet:states}. For weak stiffness constants, and for large $\nu_{\bmq}$ characteristic of long-wavelength optical modes, we can treat $V$ as a perturbation and obtain asymptotically accurate expressions for the transition amplitudes from one unperturbed Floquet state to another.

We consider the case in which $V^R$ is switched on at $t=0$, and compute the transition rates from an initial (unperturbed) Floquet state $\prod_{\bmk}\ket{\phi^{-}_{\bmk}(0)}$ to a final Floquet state $\prod_{\bmk}e^{-i\theta_{\bmk}t/T}\ket{\phi^{+}_{\bmk}(t)}$. The Floquet states can be obtained from Eqs.~\ref{eq:floquet:modes}, and~\ref{eq:floquet:amplitudes}. The transition probability $\abs{\mathcal{P}(\bmq, t)}^2$ can be obtained in the lowest perturbative order from the equation below.
\begin{equation}
    \mathcal{P}(\bmq, t)  \approx  \frac{\pi}{\omega}
    \sum_{\bmk, l , m} \frac{e^{-i\left(\theta_{\bmk}-m\pi\right)\omega t/\pi}-1}{\theta_{\bmk}-m\pi}\;V_{\bmk.\bmq}^{l l+m}.
    \label{eq:aqt}
\end{equation}
In order to obtain this expression, Floquet Perturbation Theory was used in a manner similar to ~\citep{PhysRevA.91.033601, Sen_2021}. Here, the contribution to the Dyson series of the propagator in the interaction picture was approximated by the reduced interaction after it was time-translated with the unperturbed propagator. The unperturbed propagator was spectrally decomposed into its components in the Floquet eigenbasis, and the orthonormality of the Floquet states was used to simplify the double sum. In addition, the following quantity was defined.
\begin{equation}
    V_{\bmk}^{m n}(\bmq)\equiv\mel{\phi^+_{m\bmk}}{V^R_{\bmk}(\bmq)}{\phi^-_{n\bmk}},
    \label{eq:vkq:melem}
\end{equation}
where $\ket{\phi^+_{m\bmk}}$ are the components of the Fourier series $\ket{\phi^\pm_{\bmk}(t)} = \displaystyle\sum_n \ket{\phi^{\pm}_{n\bmk}}\;e^{in\omega t}$. Let us now define the Fourier amplitudes $u^\pm_{n\bmk}(\omega), v^\pm_{n\bmk}(\omega)$, such that
\begin{align}
    |\phi^\pm_{n\bmk} \rangle &\equiv \frac{1}{T} \int^{T}_{0}dt\; |\phi^\pm_{\bmk}(t)\rangle e^{-in\omega t} \nonumber \\
    &= \frac{1}{T} \int^{T}_{0}dt\; \left[u_{\bmk}^{\pm}(t)\ket{0} + v_{\bmk}^{\pm}(t) \ket{\bmk, -\bmk} \right] \nonumber\\
    &\equiv u_{n\bmk}^{\pm}(\omega)\ket{0} + v_{n\bmk}^{\pm}(\omega)\ket{\bmk, -\bmk}. 
    \label{eq:floquet:fourier:ampl}
\end{align}
Now, noting that the reduced interaction $V^R_{\bmk,\bmq}$ in equation~\ref{eq:reduced:interaction} is anti-normal-ordered, substituting equation~\ref{eq:floquet:fourier:ampl} into equation~\ref{eq:vkq:melem}, and simplifying
the matrix element by using Wick-contractions yields
\begin{align}
    V^{mn}_{\bmk}(\bmq) &= u^{+\ast}_{m\bmk}(\omega)\;u^{-}_{n\bmk}(\omega)\;\mel{0}{V^R_{\bmk}(\bmq)}{0}\nonumber\\
    &= - \frac{|\lambda_0|^2}{\nu_{\bmq}}\;\left(\delta^{\;}_{\bmk,\frac{\bmq}{2}} + \delta^{\;}_{\bmk,-\frac{\bmq}{2}}  -2 \right)\;u_{n\bmk}^{+\ast}(\omega)\;u_{m,\bmk}^{-}(\omega)
    \label{eq:vmnkq}
\end{align}
In equation~\ref{eq:vmnkq},  we can ignore the contribution of the last term in the brackets in the RHS, since it will only contribute a time-series to the transition amplitude that has a fundamental frequency of $\omega\;,\forall\;\bmq$. Thus, it just contributes an overall constant when the system is strobed at integer multiples of $T$. Substituting equation~\ref{eq:vmnkq} into equation~\ref{eq:aqt} yields the following result.
\begin{multline}
    \mathcal{P}(\bmq, t) \sim  \sum_{m . n} \frac{e^{i\left[\theta_{\frac{\bmq}{2}}+\left(n-m\right)\pi\right]\omega t/\pi}-1}{\theta_\frac{\bmq}{2}+\left(n-m\right)\pi}\; u_{n,\frac{\bmq}{2}}^{+\ast}(\omega)\; u_{m,\frac{\bmq}{2}}^{-}(\omega).  
    \label{eq:transition:ampl}
\end{multline}
Note that we have simplified the expression in the equation above by dropping overall constant amplitudes, since they can be absorbed into arbitrary units. 
Inspection of the exponents in equation~\ref{eq:transition:ampl} reveals that, if $\theta_{\frac{\bmq}{2}}=\pi/M$ with integer $M$, the RHS is a Fourier series with fundamental mode $\Omega = \omega/M$. In particular, along the manifold given by $\bmq=2\bmk_0$, where $\bmk_0$ is the time crystal momentum obtained from equation~\ref{eq:tc:ks}, we have $\Omega = \omega/2$, an exact period-doubled subharmonic.

\subsection{Finite Size Melting of the Time Crystal}
\label{subsec:melting}
\begin{figure}[t!]
    \begin{center}
        \includegraphics[width=0.5\textwidth, keepaspectratio]{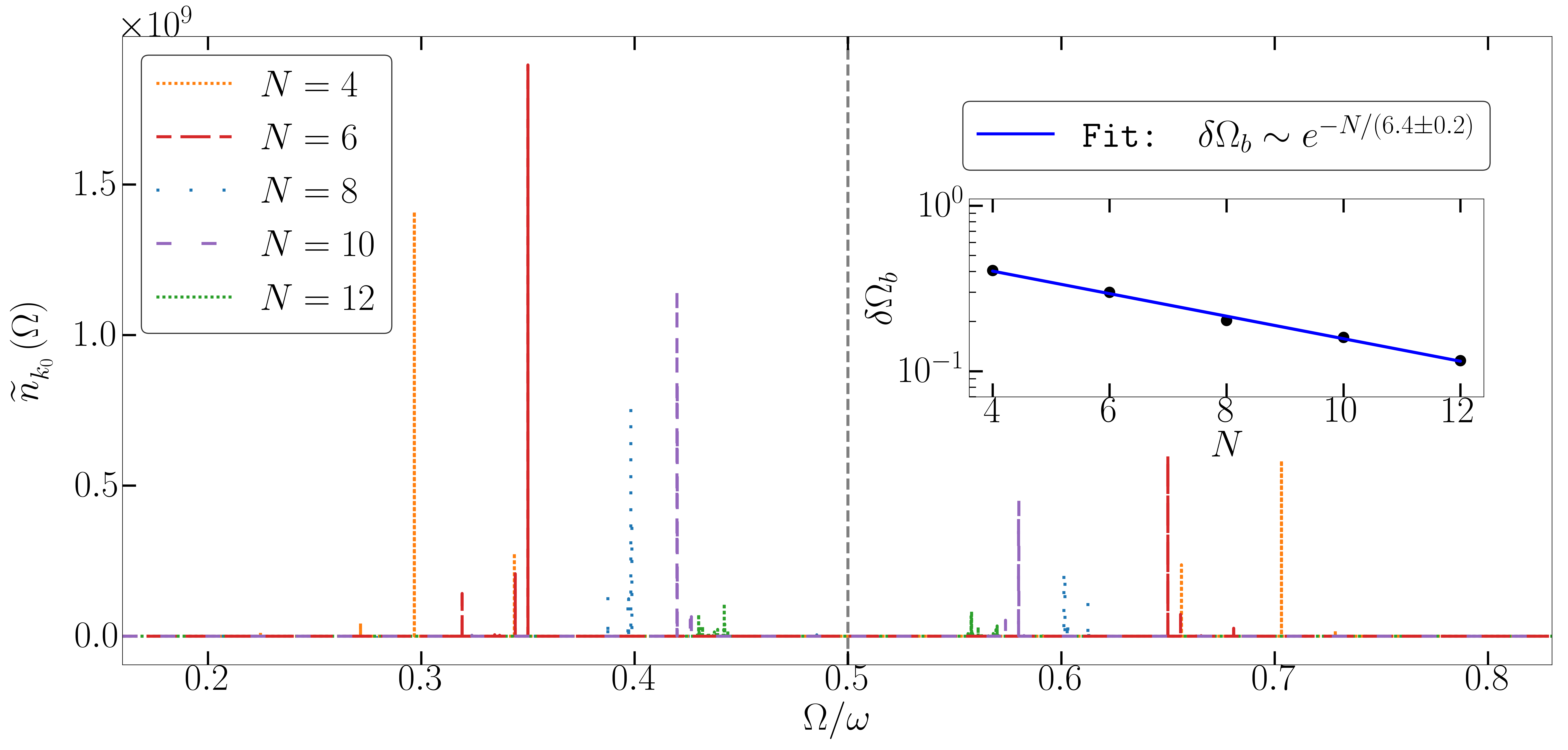}    
    \end{center}
    \caption{Dynamics of the SOC-coupled vibrating DTC in the 1D TFIM, starting from fermionic vacuum $\ket{0}$ (the fully $z-$polarized state) at $t=0$. The FFT amplitude of the fermion number at momentum $k_0$ is plotted (in arbitrary units) against the frequency domain $\Omega$ (in units of drive frequency $\omega$, obtained by solving equations~\ref{eq:tc:ks} with $k_0\approx\pi/2$) for different system sizes $N$. The expected peak position due to the subharmonic is indicated by a vertical black dashed line at $\Omega=\omega/2$. The actual peaks cluster around this value, generating beats in time evolution. The inset shows a semilog plot of the beat frequency $\delta\Omega_b$ (derived from estimates of the split between the most prominent peaks on either side of $\Omega=\omega/2$) versus $N$. The exponential decay, inferred from the curve fitting that results in $\chi^2\approx 0.004$, is consistent with the thermodynamic suppression of melting.}
    \label{fig:finitesize:melting}
\end{figure}

So far, we have shown that TTSB in the unperturbed system is not a transient, but a persistent phenomenon. To confirm that the subharmonic signal, a defining feature of this DTC, can endure disruptions from finite-sized effects of the SOC, it is vital to ensure that the timescale of any heating process, resulting from the perturbative splitting of Floquet quasienergy levels, increases with system size, ultimately disappearing in the thermodynamic limit~\citep{PhysRevLett.117.090402, melting:khemani}.

This necessitates reversing the order of limits in the calculation of the transition amplitudes (equation~\ref{eq:transition:ampl}) from infinite size then time to infinite time then size~\citep{melting:khemani,melt2023}. However, exact quantum dynamics simulations of finite-sized systems are challenging due to the NP-hard problem posed by the SOC coupling disrupting the particle-hole symmetry of the unperturbed system, leading to dynamics beyond free fermions. Yet, for the 1D case, simulations with smaller sizes are feasible on High Performance Computing (HPC) systems with Graphical Processing Units (GPUs). In our case, we first chose a $k_0$ value that falls exactly in the reciprocal lattice and simulated the dynamics with the corresponding $\omega$ value obtained by solving the equations~\ref{eq:tc:ks}. We selected a $k_0$ value as close to $\pi/2$ as permitted by the lattice resolution, keeping $\omega$ as close to the critical value $\omega_c = 2$ as possible. The functions $b_k, \Delta_k$ were chosen as $\cos{k}, \sin{k}$, respectively, consistent with the 1D TFIM system.  Using the OpenFermion package~\citep{openfermion}, matrix representations for the Hamiltonian, as well as the propagators at the end of the first and second duty cycles of the drive, were generated in the occupation number representation of the free fermions. The perturbation strength was set to ensure that the system operated within the weak coupling regime, where the phonon energies are significantly lower than the driving frequency $\omega$~\citep{floquet:pert}. The system's many-body state, $\ket{\Psi(t)}$, was evolved from vacuum in $T/2$ time increments using CuPy~\citep{cupy_learningsys2017}. The diagnosis of the DTC was achieved by evaluating $\widetilde{n}_{k_0}(\Omega)$, the Fast Fourier Transform (FFT) amplitude of $n^{\;}_{k_0}(t)=\expval{c^\dagger_{k_0}c^{\;}_{k_0}}{\Psi(t)}$, calculated using NumPy-FFT~\citep{harris2020array}.

The results, shown in figure~\ref{fig:finitesize:melting}, present the FFT amplitudes in the frequency domain ($\Omega$) for different system sizes ($N$). A distinct peak at $\Omega=\omega/2$ is expected, representing a subharmonic response at $q=2k_0$. However, for finite sizes, this peak splits into clusters on either side of the expected position. This characteristic, resulting from the phonon-induced quasienergy splitting of degenerate Floquet states, generates beats in the subharmonic, leading to the DTC's melting, similar to Stark-MBL time crystals~\citep{melt2023}. Increasing the system size brings the split peaks closer to each other and to $\Omega=\omega/2$, prolongs the beat times, and maintains the DTC's coherence for longer durations. The inset graph in figure~\ref{fig:finitesize:melting} shows the calculated beat frequency $\delta\Omega_b$ versus $N$, revealing an exponential decay pattern. Since the melting time $T_b\sim (\delta\Omega_b)^{-1}$, this aligns with reports of exponential increase in melting times for DTCs supported by MBL~\citep{PhysRevLett.117.090402, melting:khemani, melt2019, melt2020, melt2023}, suggesting that the DTC is robust against finite-size effects and that the melting process is thermodynamically resisted.

In the next section, we use the ideas discussed and results obtained so far to chart out the vibrating-time crystal phase in a specific spin-liquid described by the Kitaev model.

\section{Time Crystal in the Kitaev Spin-Liquid}
\label{sec:kitaev}
When dealing with specific cases, we have chosen to evaluate the Fourier amplitudes $u_{n,\pm\frac{\bmq}{2}}^{\pm}(\omega)$ in equation~\ref{eq:transition:ampl} by numerically sampling the time series for the Floquet states in Eqs.~\ref{eq:floquet:modes} and~\ref{eq:floquet:amplitudes} and performing FFT of the data. Numerical errors can be rendered irrelevant for subharmonic modes, as long as the sample size is large enough and the Nyquist rate of the sampling is substantially greater than the drive frequency $\omega$~\footnote{See, for example,~\citep{Press2007}}.

The Kitaev model on a brick lattice~\citep{Chen_2008} is described by the Hamiltonian
\begin{align}
    H_{\mathrm{K}}\left[J_{1}, J_{2}, J_{3}\right] &= \sum_{\left\langle\bmr \bmr^{\prime}\right\rangle ; \text { xlink }} J_{1} \sigma_{\mathbf{r}}^{x} \sigma_{{\bmr}^{\prime}}^{x} + \sum_{\left\langle\bmr \bmr^{\prime}\right\rangle ; \text { ylink }} J_{2} \sigma_{\mathbf{r}}^{y} \sigma_{\mathbf{r}^{\prime}}^{y} \nonumber \\
    &+ \sum_{\left\langle{\bmr} \boldsymbol{r}^{\prime}\right\rangle ; \text { zlink }} J_{3} \sigma_{\mathbf{r}}^{z} \sigma_{\mathbf{r}^{\prime}}^{z},
    \label{eq:kitaev}
\end{align}
\begin{figure}[t!]
    \begin{center}
        \includegraphics[width=0.48\textwidth, keepaspectratio]{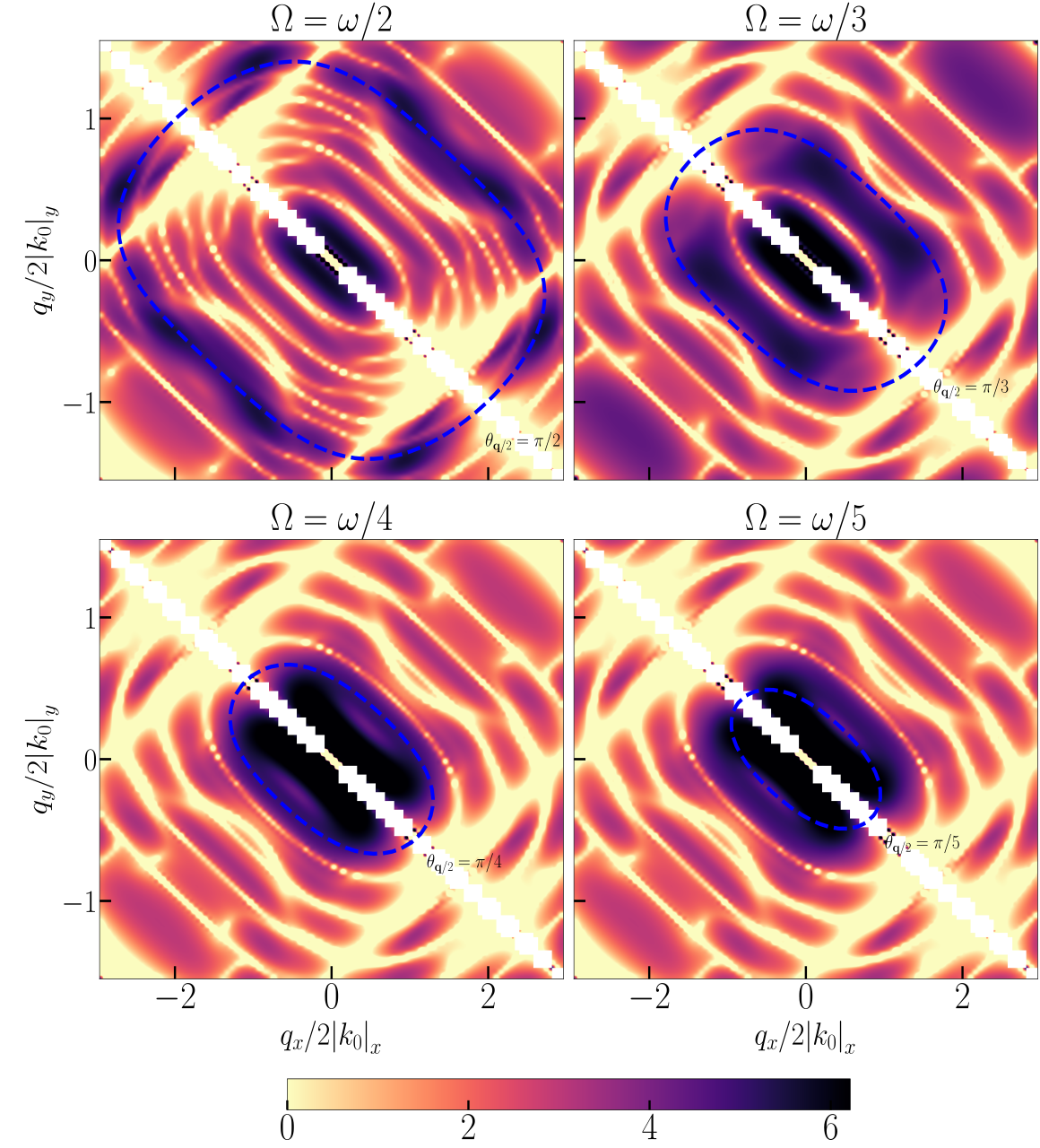}    
    \end{center}
    \caption{Phonon FFT cross-sections at specific frequencies $\Omega$ for a periodically driven two-dimensional fermionic system of size $100\times 100$, described by the Hamiltonian in equation~\ref{eq:general:hamiltonian}. Each cross section displays $\norm{\tilde{\mathcal{P}}(\Omega, \bmq)}$, the FFT of the time series data obtained from $\norm{\mathcal{P}(\Omega, \bmq)}$ in~\ref{eq:transition:ampl}, as a function of the phonon momenta $\bmq=(q_x, q_y)$, for a particular value of $\Omega/\omega$, where $\omega$ is the drive frequency. Here, $g_0$ and $\omega$ are chosen in such a way that equation~\ref{eq:tc:ks} is satisfied for the fermion momentum $\bm{k}_0\approx \big(\pi/6, \pi/3\big)$. We assume an isotropic energy dispersion ($\alpha=1$ in equation~\ref{eq:kitaev:tck}) with $g_1=1.0$. The dashed blue curve in each panel represents the manifold $\theta_{\frac{\bm{q}}{2}}={\pi}/{M}$, with $M=\omega/\Omega$.}
    \label{fig:2dscattampl}
\end{figure}
where $\sigma^{\alpha}$ for $\alpha=x, y, z$ indicate Pauli matrices and $\left\langle\mathbf{r r}^{\prime}\right\rangle$ denote neighboring site indices on the brick lattice and $J_{i}$, for $i=1,2,3$ denotes the magnitude of the interaction between neighboring spins. The link in the brick lattice can be of types $x, y$ and $z$ and these are, as shown in~\citep{Chen_2008}, roughly analogous to the right, left, and top bonds of the more well-known Kitaev honeycomb model, which hosts spin interactions $x-x, y-y$ and $z-z$, respectively. A straightforward analysis shows that this model, in the ground state sector, can be described by $H$ in equation~\ref{eq:general:hamiltonian} with the following identification.
\begin{equation}
    g_{0}=J_{3}, \; b_{\bmk}=J_{1}\left(\cos k_{x}+\alpha \cos k_{y}\right), \; \Delta_{\bmk}=J_{1}\left(\sin k_{x}+\alpha \sin k_{y}\right)
    \label{eq:kitaev:tck}
\end{equation}
where $\alpha=J_{2} / J_{1}$ is the anisotropy parameter. In what follows, we scale all energy and frequency scales by $J_{1}$. Our driving protocol would be two square waves with $50\%$ duty cycle, one that flips the value of $J_1$ from $1$ to $0$ and the other that flips the value of $J_3$ from $g_0$ to $g_1$. In principle, the second square wave can be forgone simply by setting $g_0=g_1$. Setting $J_2$ to $0$ recovers the one-dimensional case.

With the identification of the Kitaev model, the analysis in Subsection~\ref{subsec:subharmonics} predicts that there will be a stable time crystal phase at the point in the Brillouin zone given by equations~\ref{eq:tcpoint:kitaev}. Note that for any finite $0<\alpha\leq 1$, the time crystal phase is robust against changes of $J_{3}$ and $\alpha$; such changes merely change the value of $k_{x}$ and $k_{y}$ at which equation~\ref{eq:tcpoint:kitaev} is satisfied. The stability of the phase disappears at $\alpha=0$, for which the model is reduced to an effective 1D model. Also, for the Kitaev model, spin-correlators like $\mel{\psi(nT)}{\sum_{\left\langle\mathbf{r r}^{\prime}\right\rangle} \sigma_{\mathbf{r}}^{x} \sigma_{\mathbf{r}^{\prime}}^{x}}{\psi(nT)}$,  can be directly mapped to fermionic bilinear operators~\citep{Chen_2008}. To detect subharmonics in the vibrational modes of this system, we chose the range of phonon momenta as $q_x, q_y\in \big[-\pi, \pi\big]$ for simplicity. The various panels in figure~\ref{fig:2dscattampl} show visual representations of $|\widetilde{\mathcal{P}}_{\bmq}(\Omega)|$, the FFT of the transition rate $|\mathcal{P}(\bmq, t)|$ as obtained from equation~\ref{eq:aqt}, for different $\Omega$, in the $q_x, q_y$ plane. One can clearly see very stable subharmonics along the manifold $\theta_{\frac{\bm{q}}{2}}=\displaystyle\frac{\pi}{M}$ when $\Omega=\omega/M$. Thus, the subharmonic response is easily recognized in the dynamics of the phonon modes in such systems, overcoming the realization challenge described in Section~\ref{sec:stability:detection}.

Upon examining the points along the dashed manifold in figure~\ref{fig:2dscattampl}, the DTC phase can be clearly detected for most phonon frequencies in its immediate vicinity. Thus, this phase exhibits a high degree of rigidity in response to variations in drive parameters. However, DTC melts in the region where the dashed manifold meets the condition $k_x\approx-k_y$, since equation \ref{eq:subh:melts} is met.

\begin{figure}[t!]
    \includegraphics[width=0.48\textwidth, keepaspectratio]{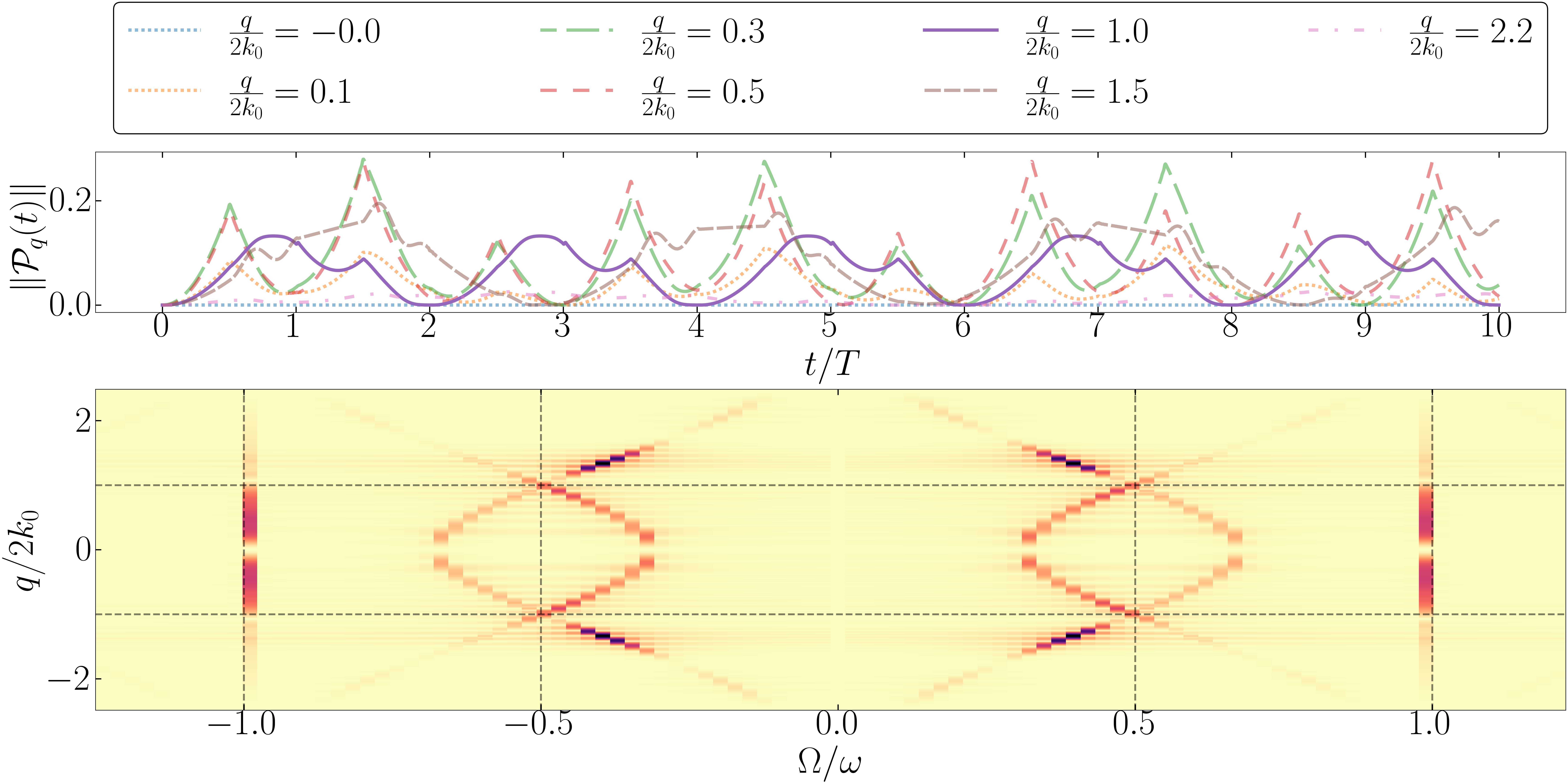}
    \caption{Phonon Fourier modes and time-series plots for the 1D case with $1000$ fermions. Here, $b_k = \cos{k}, \Delta_k=\sin{k}$, consistent with the 1D TFIM. The values of amplitude $g_0$ and drive frequency $\omega$ are chosen so that the equation~\ref{eq:tc:ks} is satisfied for the momentum ${k}_0\approx 0.632092$ and $ g_1=1.0$. The top panel shows the scattering amplitude $\abs{\mathcal{P}(q, t)}$ over time for different values of the phonon momentum $q$ (in units of $2k_0$). The bottom panel visualizes $|\widetilde{\mathcal{P}}_{\bmq}(\Omega)|$, the FFTs of the transition rate $|\mathcal{P}(q, t)|$ as obtained from equation~\ref{eq:aqt}. The subharmonic signal can be seen in panels when $q=2k_0$.}
\label{fig:1dscattampl}
\end{figure}

We have also investigated the vibrating time crystal for the special case where $\alpha=0$, when it reduces to the 1D TFIM. Figure~\ref{fig:1dscattampl} plots time series data (top panel) of $|\mathcal{P}(q, t)|$ for various values of $q$, as well as FFTs of the same, plotted against $q, \Omega$. Clearly, when $q=2k_0$, the FFT has a peak at $\Omega=\pm\omega/2$. Additional peaks are seen at $\pm\omega/3, \pm\omega/4, \pm\omega/5 \dots$ corresponding to additional subharmonics. The instability of this subharmonic can easily be seen, as slight variations of $g_0$ (and consequentially $k_0$) will take the system out of a subharmonic and between two subharmonics.
\subsection{Time Crystal Phase Diagram}
\label{subsec:phasediag}
From equation.~\ref{eq:transition:ampl}, it is evident that the time crystal may melt even along the manifold given by $\theta_{\frac{\bmq}{2}}=\pi/M$ if the amplitude of the fundamental mode vanishes. This will occur in the submanifold given by $f_{\omega}(\bmq)=0$, where
\begin{equation}
f_{\omega}(\bmq) \equiv \sum_n u_{n,\frac{\bmq}{2}}^{+\ast}(\omega)\; u_{n,\frac{\bmq}{2}}^{-}(\omega).
\label{eq:subh:melts}
\end{equation}
In that case, the terms involving $n-m=\pm 1, \pm 2 \dots$ continue to contribute to the RHS of equation~\ref{eq:transition:ampl}, and the response may have a longer period, thus causing the time crystal to melt. However, higher-order terms in $\mathcal{P}(\bmq, t)$ from Floquet Perturbation Theory do not disrupt the subharmonic in the weak coupling limit, consistent with the finite-size numerics in section~\ref{subsec:melting}. This happens because the subharmonics seen in the Floquet perturbation theory arises from the phase components, such as those in equation~\ref{eq:transition:ampl}, which persist at higher orders~\citep{PhysRevA.91.033601}. Melting can still occur due to other factors beyond the scope of this particular model, such as defects and disorder.
In addition to this, the subharmonic phase will disappear into a Floquet spin-liquid phase when equations~\ref{eq:tcpoint:kitaev} no longer allow real roots for $\bmk_0=\left(k^x_0, k^y_0\right)$. For the case where $n=1$, the equations can be combined to yield
\begin{equation}
\cos(k^x_0 - k^y_0) = \frac{1}{2\alpha}\left(g^2_0 + \frac{\omega^2}{4}-\alpha^2-1\right).
\end{equation}
Therefore, a subharmonic will only be possible for a particular pair $\omega, g_0$ as long as $R^2_-\leq4g^2_0+\omega^2\leq R_+^2$, where $R_\pm = 2\left(1\pm\alpha\right)$. The phase profile thus consists of two phases, a temporally ordered time crystal and a Floquet spin-liquid state. The manifold of transition points that separate these phases is the boundary of the intersection between two circles of radius $R_\pm$. 
\begin{figure}[t!]
    \centering
    \includegraphics[width=0.45\textwidth, keepaspectratio]{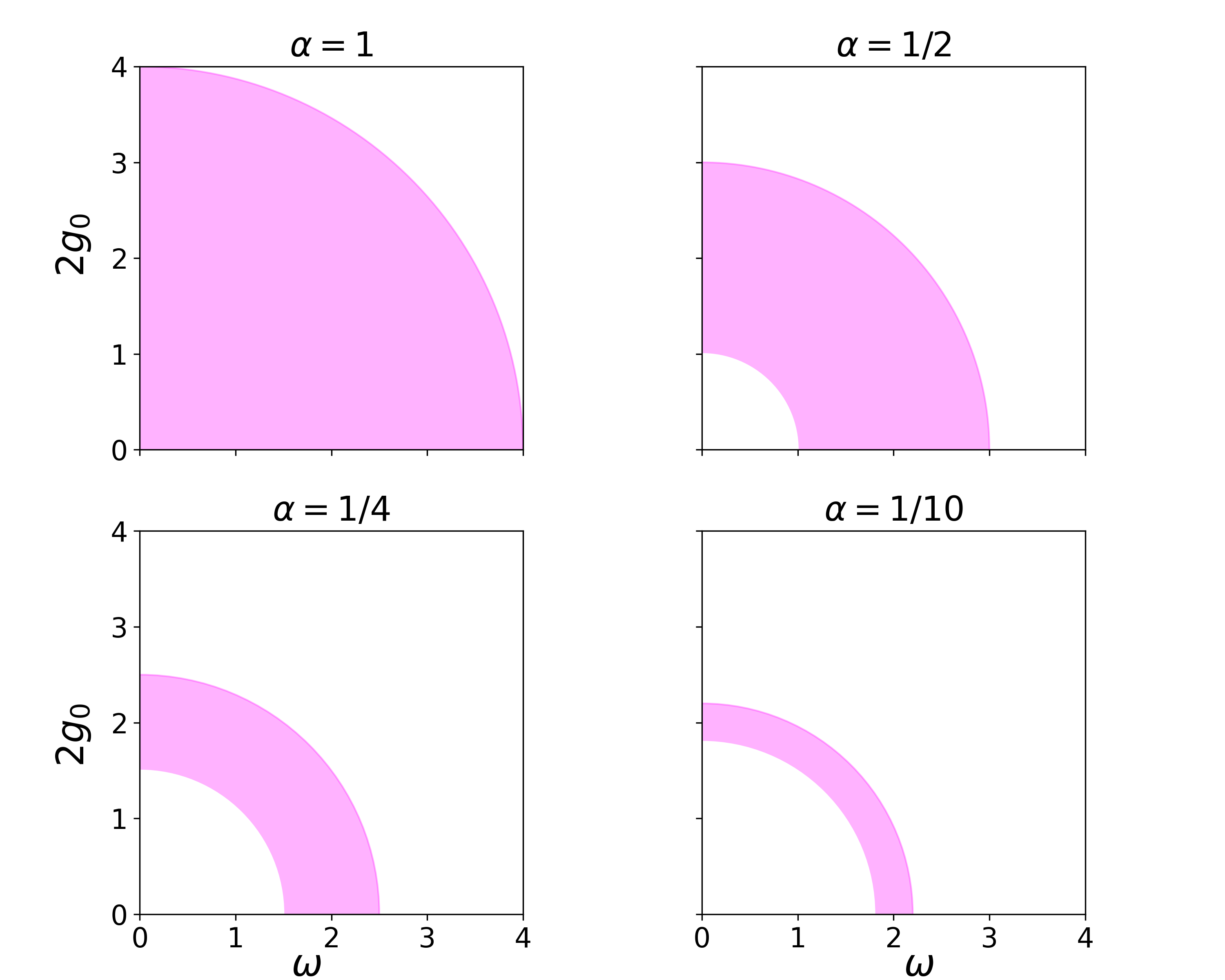}
    \caption{Phase diagram for the time crystal phase (represented by the magenta-colored region) and disordered Floquet spin-liquid  phase (white region) in the $\left(2g_{0},\omega\right)$-plane for several values of $\alpha$. The phase boundary shrinks to a point at $\alpha=0$ where the model becomes $1 \mathrm{D}$ (since $J_{2}=0$ at this point).}
    \label{fig:phasediag}
\end{figure}
The phase diagram in the $\omega, g_0$ plane is shown for several representative values of $\alpha$ in figure~\ref{fig:phasediag}. As was originally the case in the unperturbed system,  the limit $J_2\rightarrow 0$ recovers the $1$ dimensional case, where the manifold converges to a point.

\section{Conclusions and Outlook}
To summarize, our investigation focused on the generation of Floquet-time crystals in higher dimensions through the time-periodic driving of integrable free-fermionic models. By utilizing fermion-boson coupling models like magnetoelastic spin-orbit coupling, we were able to achieve a robust time-crystal phase that can be detected using novel techniques. This approach allows for the realization of time crystals without relying on many-body localization, ensuring long-term stability and great potential for applications in quantum technology. Our research also emphasized the significance of studying subharmonic responses and their implementation in a Kitaev spin liquid, which contributes to our understanding of time translational symmetry breaking and its practical implications. Further investigations could involve examining the scaling laws for universality near the critical point, specifically when the anisotropy $\alpha$ tends to zero in the Kitaev spin liquid. Moreover, there is the possibility of experimentally realizing this time crystal in optical lattices and cavity radiation, which can be theoretically modeled using Dicke or Jaynes-Cummings coupling. Exploring the melting dynamics of the vibrating DTC beyond the weak coupling limit is also a possibility. Lastly, exploring the observation of time crystals in other integrable models, such as the massive Thirring model, Tonks gases, or generalized Gaudin models, which can be solved using Bethe Ansatz, would significantly expand the range of experimental candidates for realizing the time-crystal phenomenon.

\section*{CRediT author statement}
\textbf{Rahul Chandra:} Formal analysis, Visualization, Writing Original draft preparation. \textbf{Analabha Roy}  Conceptualization, Methodology, Software, Data Curation, Validation, Resources, Writing-Reviewing and Editing, Funding acquisition.

\section*{Acknowledgements}
 Both authors thank Prof. Krishnendu Sengupta, School of Physical Sciences, Indian Association for the Cultivation of Science, Kolkata, for supervision and guidance.
 
Funding: This work was supported by the Science and Engineering Research Board (SERB, Govt. of India) and the University Grants Commission (UGC, Govt. of India) [Grant Numbers: TAR/2018/000080, CRG/2018/004002, 1407/CSIR-UGC NET JUNE 2018].

\bibliographystyle{elsarticle-num}
\bibliography{references}

\begin{thebibliography}{10}
\expandafter\ifx\csname url\endcsname\relax
  \def\url#1{\texttt{#1}}\fi
\expandafter\ifx\csname urlprefix\endcsname\relax\def\urlprefix{URL }\fi
\expandafter\ifx\csname href\endcsname\relax
  \def\href#1#2{#2} \def\path#1{#1}\fi

\bibitem{PhysRevLett.109.160401}
F.~Wilczek, Quantum time crystals, Phys. Rev. Lett. 109 (2012) 160401.
\newblock \href {https://doi.org/10.1103/PhysRevLett.109.160401}
  {\path{doi:10.1103/PhysRevLett.109.160401}}.

\bibitem{PhysRevA.91.033617}
K.~Sacha, Modeling spontaneous breaking of time-translation symmetry, Phys.
  Rev. A 91 (2015) 033617.
\newblock \href {https://doi.org/10.1103/PhysRevA.91.033617}
  {\path{doi:10.1103/PhysRevA.91.033617}}.

\bibitem{PhysRevLett.116.250401}
V.~Khemani, A.~Lazarides, R.~Moessner, S.~L. Sondhi, Phase structure of driven
  quantum systems, Phys. Rev. Lett. 116 (2016) 250401.
\newblock \href {https://doi.org/10.1103/PhysRevLett.116.250401}
  {\path{doi:10.1103/PhysRevLett.116.250401}}.

\bibitem{PhysRevLett.118.030401}
N.~Y. Yao, A.~C. Potter, I.-D. Potirniche, A.~Vishwanath, Discrete time
  crystals: Rigidity, criticality, and realizations, Phys. Rev. Lett. 118
  (2017) 030401.
\newblock \href {https://doi.org/10.1103/PhysRevLett.118.030401}
  {\path{doi:10.1103/PhysRevLett.118.030401}}.

\bibitem{PhysRevLett.111.070402}
P.~Bruno, Impossibility of spontaneously rotating time crystals: A no-go
  theorem, Phys. Rev. Lett. 111 (2013) 070402.
\newblock \href {https://doi.org/10.1103/PhysRevLett.111.070402}
  {\path{doi:10.1103/PhysRevLett.111.070402}}.

\bibitem{PhysRevLett.114.251603}
H.~Watanabe, M.~Oshikawa, Absence of quantum time crystals, Phys. Rev. Lett.
  114 (2015) 251603.
\newblock \href {https://doi.org/10.1103/PhysRevLett.114.251603}
  {\path{doi:10.1103/PhysRevLett.114.251603}}.

\bibitem{PhysRevLett.117.090402}
D.~V. Else, B.~Bauer, C.~Nayak, Floquet time crystals, Phys. Rev. Lett. 117
  (2016) 090402.
\newblock \href {https://doi.org/10.1103/PhysRevLett.117.090402}
  {\path{doi:10.1103/PhysRevLett.117.090402}}.

\bibitem{10.1063/PT.3.4020}
N.~Y. Yao, C.~Nayak, {Time crystals in periodically driven systems}, Physics
  Today 71~(9) (2018) 40--47.
\newblock \href {https://doi.org/10.1063/PT.3.4020}
  {\path{doi:10.1063/PT.3.4020}}.

\bibitem{ermodel:emch}
G.~G. Emch, {Non‐Markovian Model for the Approach to Equilibrium}, Journal of
  Mathematical Physics 7~(7) (1966) 1198--1206.
\newblock \href {https://doi.org/10.1063/1.1705023}
  {\path{doi:10.1063/1.1705023}}.

\bibitem{ermodel:radin}
C.~Radin, {Approach to Equilibrium in a Simple Model}, Journal of Mathematical
  Physics 11~(10) (1970) 2945--2955.
\newblock \href {https://doi.org/10.1063/1.1665079}
  {\path{doi:10.1063/1.1665079}}.

\bibitem{PhysRevB.94.085112}
C.~W. von Keyserlingk, V.~Khemani, S.~L. Sondhi, Absolute stability and
  spatiotemporal long-range order in floquet systems, Phys. Rev. B 94 (2016)
  085112.
\newblock \href {https://doi.org/10.1103/PhysRevB.94.085112}
  {\path{doi:10.1103/PhysRevB.94.085112}}.

\bibitem{huse.mbl}
R.~Nandkishore, D.~Huse, Many-body localization and thermalization in quantum
  statistical mechanics, Annual Review of Condensed Matter Physics 6~(1) (2015)
  15--38.
\newblock \href {https://doi.org/10.1146/annurev-conmatphys-031214-014726}
  {\path{doi:10.1146/annurev-conmatphys-031214-014726}}.

\bibitem{PhysRevResearch.1.033202}
F.~Machado, G.~D. Kahanamoku-Meyer, D.~V. Else, C.~Nayak, N.~Y. Yao,
  Exponentially slow heating in short and long-range interacting floquet
  systems, Phys. Rev. Res. 1 (2019) 033202.
\newblock \href {https://doi.org/10.1103/PhysRevResearch.1.033202}
  {\path{doi:10.1103/PhysRevResearch.1.033202}}.

\bibitem{PhysRevX.7.011026}
D.~V. Else, B.~Bauer, C.~Nayak, Prethermal phases of matter protected by
  time-translation symmetry, Phys. Rev. X 7 (2017) 011026.
\newblock \href {https://doi.org/10.1103/PhysRevX.7.011026}
  {\path{doi:10.1103/PhysRevX.7.011026}}.

\bibitem{Mori_2018}
T.~Mori, T.~N. Ikeda, E.~Kaminishi, M.~Ueda, Thermalization and
  prethermalization in isolated quantum systems: a theoretical overview,
  Journal of Physics B: Atomic, Molecular and Optical Physics 51~(11) (2018)
  112001.
\newblock \href {https://doi.org/10.1088/1361-6455/aabcdf}
  {\path{doi:10.1088/1361-6455/aabcdf}}.

\bibitem{PhysRevLett.129.140602}
K.~Bull, A.~Hallam, Z.~Papi\ifmmode~\acute{c}\else \'{c}\fi{}, I.~Martin,
  Tuning between continuous time crystals and many-body scars in long-range
  $xyz$ spin chains, Phys. Rev. Lett. 129 (2022) 140602.
\newblock \href {https://doi.org/10.1103/PhysRevLett.129.140602}
  {\path{doi:10.1103/PhysRevLett.129.140602}}.

\bibitem{Pizzi2021}
A.~Pizzi, J.~Knolle, A.~Nunnenkamp, Higher-order and fractional discrete time
  crystals in clean long-range interacting systems, Nature Communications
  12~(1) (2021) 2341.
\newblock \href {https://doi.org/10.1038/s41467-021-22583-5}
  {\path{doi:10.1038/s41467-021-22583-5}}.

\bibitem{Zhang2018}
X.~Zhang, K.~Zhang, Y.~Shen, S.~Zhang, J.-N. Zhang, M.-H. Yung, J.~Casanova,
  J.~S. Pedernales, L.~Lamata, E.~Solano, K.~Kim, Experimental quantum
  simulation of fermion-antifermion scattering via boson exchange in a trapped
  ion, Nature Communications 9~(1) (2018) 195.
\newblock \href {https://doi.org/10.1038/s41467-017-02507-y}
  {\path{doi:10.1038/s41467-017-02507-y}}.

\bibitem{mbeng_quantum_2020}
G.~B. Mbeng, A.~Russomanno, G.~E. Santoro, The quantum {Ising} chain for
  beginnersArXiv:2009.09208 [cond-mat, physics:quant-ph] (sep 2020).
\newblock \href {https://doi.org/10.48550/arXiv.2009.09208}
  {\path{doi:10.48550/arXiv.2009.09208}}.

\bibitem{Chen_2008}
H.-D. Chen, Z.~Nussinov, Exact results of the kitaev model on a hexagonal
  lattice: spin states, string and brane correlators, and anyonic excitations,
  Journal of Physics A: Mathematical and Theoretical 41~(7) (2008) 075001.
\newblock \href {https://doi.org/10.1088/1751-8113/41/7/075001}
  {\path{doi:10.1088/1751-8113/41/7/075001}}.

\bibitem{soc:dixon}
G.~S. Dixon, Lattice thermal conductivity of antiferromagnetic insulators,
  Phys. Rev. B 21 (1980) 2851--2864.
\newblock \href {https://doi.org/10.1103/PhysRevB.21.2851}
  {\path{doi:10.1103/PhysRevB.21.2851}}.

\bibitem{fiete:phonons}
G.~L. Stamokostas, P.~E. Lapas, G.~A. Fiete, Thermal conductivity of local
  moment models with strong spin-orbit coupling, Phys. Rev. B 95 (2017) 064410.
\newblock \href {https://doi.org/10.1103/PhysRevB.95.064410}
  {\path{doi:10.1103/PhysRevB.95.064410}}.

\bibitem{doppler:phonons}
B.~Xia, Z.~Jiang, L.~Tong, S.~Zheng, X.~Man, Topological bound states in
  elastic phononic plates induced by disclinations, Acta Mechanica Sinica
  38~(2) (2022) 521459.
\newblock \href {https://doi.org/10.1007/s10409-021-09083-0}
  {\path{doi:10.1007/s10409-021-09083-0}}.

\bibitem{Jahng2023}
J.~Jahng, S.~Lee, S.-G. Hong, C.~J. Lee, S.~G. Menabde, M.~S. Jang, D.-H. Kim,
  J.~Son, E.~S. Lee, Characterizing and controlling infrared phonon anomaly of
  bilayer graphene in optical-electrical force nanoscopy, Light: Science {\&}
  Applications 12~(1) (2023) 281.
\newblock \href {https://doi.org/10.1038/s41377-023-01320-1}
  {\path{doi:10.1038/s41377-023-01320-1}}.

\bibitem{Gadre2022}
C.~A. Gadre, X.~Yan, Q.~Song, J.~Li, L.~Gu, H.~Huyan, T.~Aoki, S.-W. Lee,
  G.~Chen, R.~Wu, X.~Pan, Nanoscale imaging of phonon dynamics by electron
  microscopy, Nature 606~(7913) (2022) 292--297.
\newblock \href {https://doi.org/10.1038/s41586-022-04736-8}
  {\path{doi:10.1038/s41586-022-04736-8}}.

\bibitem{matsuda_fundamentals_2015}
O.~Matsuda, M.~C. Larciprete, R.~L. Voti, O.~B. Wright, Fundamentals of
  picosecond laser ultrasonics, Ultrasonics 56 (2015) 3--20.
\newblock \href {https://doi.org/10.1016/j.ultras.2014.06.005}
  {\path{doi:10.1016/j.ultras.2014.06.005}}.

\bibitem{Ruello2015-qx}
P.~Ruello, V.~E. Gusev, Physical mechanisms of coherent acoustic phonons
  generation by ultrafast laser action, Ultrasonics 56 (2015) 21--35.

\bibitem{ng_excitation_2022}
R.~C. Ng, A.~El~Sachat, F.~Cespedes, M.~Poblet, G.~Madiot,
  J.~Jaramillo-Fernandez, O.~Florez, P.~Xiao, M.~Sledzinska, C.~M.
  Sotomayor-Torres, E.~Chavez-Angel, Excitation and detection of acoustic
  phonons in nanoscale systems, Nanoscale 14~(37) (2022) 13428--13451,
  publisher: The Royal Society of Chemistry.
\newblock \href {https://doi.org/10.1039/D2NR04100F}
  {\path{doi:10.1039/D2NR04100F}}.

\bibitem{Floquet}
G.~Floquet, Sur les \'equations diff\'erentielles lin\'eaires \`a coefficients
  p\'eriodiques, Annales scientifiques de l'\'Ecole Normale Sup\'erieure 2e
  s{\'e}rie, 12 (1883) 47--88.
\newblock \href {https://doi.org/10.24033/asens.220}
  {\path{doi:10.24033/asens.220}}.

\bibitem{reichl}
L.~E. Reichl, {T}he {T}ransition to {C}haos: {C}onservative {C}lassical and
  {Q}uantum {S}ystems, Springer, 2021, Ch. Time-Periodic Quantum Systems, pp.
  339--396.
\newblock \href {https://doi.org/10.1007/978-3-030-63534-3}
  {\path{doi:10.1007/978-3-030-63534-3}}.

\bibitem{Conn2000}
A.~R. Conn, N.~I.~M. Gould, P.~L. Toint, Trust Region Methods, Society for
  Industrial and Applied Mathematics, 2000.
\newblock \href {https://doi.org/10.1137/1.9780898719857}
  {\path{doi:10.1137/1.9780898719857}}.

\bibitem{2020SciPy-NMeth}
P.~Virtanen, R.~Gommers, T.~E. Oliphant, M.~Haberland, T.~Reddy, D.~Cournapeau,
  E.~Burovski, P.~Peterson, W.~Weckesser, J.~Bright, S.~J. {van der Walt},
  M.~Brett, J.~Wilson, K.~J. Millman, N.~Mayorov, A.~R.~J. Nelson, E.~Jones,
  R.~Kern, E.~Larson, C.~J. Carey, {\.I}.~Polat, Y.~Feng, E.~W. Moore,
  J.~{VanderPlas}, D.~Laxalde, J.~Perktold, R.~Cimrman, I.~Henriksen, E.~A.
  Quintero, C.~R. Harris, A.~M. Archibald, A.~H. Ribeiro, F.~Pedregosa, P.~{van
  Mulbregt}, {SciPy 1.0 Contributors}, {{SciPy} 1.0: Fundamental Algorithms for
  Scientific Computing in Python}, Nature Methods 17 (2020) 261--272.
\newblock \href {https://doi.org/10.1038/s41592-019-0686-2}
  {\path{doi:10.1038/s41592-019-0686-2}}.

\bibitem{soc:phonons}
A.~Metavitsiadis, W.~Brenig, Phonon renormalization in the kitaev quantum spin
  liquid, Phys. Rev. B 101 (2020) 035103.
\newblock \href {https://doi.org/10.1103/PhysRevB.101.035103}
  {\path{doi:10.1103/PhysRevB.101.035103}}.

\bibitem{marder:ch13}
M.~P. {M}arder, Condensed Matter Physics, John Wiley \& Sons, Ltd, 2010,
  Ch.~13, pp. 341--378.
\newblock \href {https://doi.org/10.1002/9780470949955.ch13}
  {\path{doi:10.1002/9780470949955.ch13}}.

\bibitem{PhysRevB.102.235154}
M.~Sarkar, K.~Sengupta, Dynamical transition for a class of integrable models
  coupled to a bath, Phys. Rev. B 102 (2020) 235154.
\newblock \href {https://doi.org/10.1103/PhysRevB.102.235154}
  {\path{doi:10.1103/PhysRevB.102.235154}}.

\bibitem{10.1093/acprof:oso/9780199213900.001.0001}
H.-P. Breuer, F.~Petruccione, {The Theory of Open Quantum Systems}, Oxford
  University Press, 2007.
\newblock \href {https://doi.org/10.1093/acprof:oso/9780199213900.001.0001}
  {\path{doi:10.1093/acprof:oso/9780199213900.001.0001}}.

\bibitem{PhysRevA.91.033601}
T.~Bilitewski, N.~R. Cooper, Scattering theory for floquet-bloch states, Phys.
  Rev. A 91 (2015) 033601.
\newblock \href {https://doi.org/10.1103/PhysRevA.91.033601}
  {\path{doi:10.1103/PhysRevA.91.033601}}.

\bibitem{Sen_2021}
A.~Sen, D.~Sen, K.~Sengupta, Analytic approaches to periodically driven closed
  quantum systems: methods and applications, Journal of Physics: Condensed
  Matter 33~(44) (2021) 443003.
\newblock \href {https://doi.org/10.1088/1361-648X/ac1b61}
  {\path{doi:10.1088/1361-648X/ac1b61}}.

\bibitem{melting:khemani}
V.~Khemani, R.~Moessner, S.~L. Sondhi, A brief history of time crystals (2019).
\newblock \href {http://arxiv.org/abs/1910.10745} {\path{arXiv:1910.10745}},
  \href {https://doi.org/10.48550/arXiv.1910.10745}
  {\path{doi:10.48550/arXiv.1910.10745}}.

\bibitem{melt2023}
S.~Liu, S.-X. Zhang, C.-Y. Hsieh, S.~Zhang, H.~Yao, Discrete time crystal
  enabled by {Stark} many-body localization, Physical Review Letters 130~(12)
  (2023) 120403, arXiv:2208.02866 [cond-mat, physics:quant-ph].
\newblock \href {https://doi.org/10.1103/PhysRevLett.130.120403}
  {\path{doi:10.1103/PhysRevLett.130.120403}}.

\bibitem{openfermion}
J.~R. McClean, N.~C. Rubin, K.~J. Sung, I.~D. Kivlichan, X.~Bonet-Monroig,
  Y.~Cao, C.~Dai, E.~S. Fried, C.~Gidney, B.~Gimby, P.~Gokhale, T.~H\"{a}ner,
  T.~Hardikar, V.~Havlíček, O.~Higgott, C.~Huang, J.~Izaac, Z.~Jiang, X.~Liu,
  S.~McArdle, M.~Neeley, T.~O’Brien, B.~O’Gorman, I.~Ozfidan, M.~D. Radin,
  J.~Romero, N.~P.~D. Sawaya, B.~Senjean, K.~Setia, S.~Sim, D.~S. Steiger,
  M.~Steudtner, Q.~Sun, W.~Sun, D.~Wang, F.~Zhang, R.~Babbush, Openfermion: the
  electronic structure package for quantum computers, Quantum Science and
  Technology 5~(3) (2020) 034014.
\newblock \href {https://doi.org/10.1088/2058-9565/ab8ebc}
  {\path{doi:10.1088/2058-9565/ab8ebc}}.

\bibitem{floquet:pert}
R.~Ghosh, B.~Mukherjee, K.~Sengupta, Floquet perturbation theory for
  periodically driven weakly interacting fermions, Phys. Rev. B 102 (2020)
  235114.
\newblock \href {https://doi.org/10.1103/PhysRevB.102.235114}
  {\path{doi:10.1103/PhysRevB.102.235114}}.

\bibitem{cupy_learningsys2017}
R.~Okuta, Y.~Unno, D.~Nishino, S.~Hido, C.~Loomis,
  \href{http://learningsys.org/nips17/assets/papers/paper_16.pdf}{Cupy: A
  numpy-compatible library for nvidia gpu calculations}, in: Proceedings of
  Workshop on Machine Learning Systems (LearningSys) in The Thirty-first Annual
  Conference on Neural Information Processing Systems (NIPS), 2017.
\newline\urlprefix\url{http://learningsys.org/nips17/assets/papers/paper_16.pdf}

\bibitem{harris2020array}
C.~R. Harris, K.~J. Millman, S.~J. van~der Walt, R.~Gommers, P.~Virtanen,
  D.~Cournapeau, E.~Wieser, J.~Taylor, S.~Berg, N.~J. Smith, R.~Kern, M.~Picus,
  S.~Hoyer, M.~H. van Kerkwijk, M.~Brett, A.~Haldane, J.~F. del R{\'{i}}o,
  M.~Wiebe, P.~Peterson, P.~G{\'{e}}rard-Marchant, K.~Sheppard, T.~Reddy,
  W.~Weckesser, H.~Abbasi, C.~Gohlke, T.~E. Oliphant, Array programming with
  {NumPy}, Nature 585~(7825) (2020) 357--362.
\newblock \href {https://doi.org/10.1038/s41586-020-2649-2}
  {\path{doi:10.1038/s41586-020-2649-2}}.

\bibitem{melt2019}
F.~M. Surace, A.~Russomanno, M.~Dalmonte, A.~Silva, R.~Fazio, F.~Iemini,
  Floquet time crystals in clock models, Physical Review B 99~(10) (2019)
  104303.
\newblock \href {https://doi.org/10.1103/PhysRevB.99.104303}
  {\path{doi:10.1103/PhysRevB.99.104303}}.

\bibitem{melt2020}
A.~Pizzi, D.~Malz, G.~De~Tomasi, J.~Knolle, A.~Nunnenkamp, Time crystallinity
  and finite-size effects in clean {F}loquet systems, Physical Review B
  102~(21) (2020) 214207, publisher: American Physical Society.
\newblock \href {https://doi.org/10.1103/PhysRevB.102.214207}
  {\path{doi:10.1103/PhysRevB.102.214207}}.

\bibitem{Press2007}
W.~H. Press, S.~A. Teukolsky, W.~T. Vetterling, B.~P. Flannery, Numerical
  Recipes 3rd Edition: The Art of Scientific Computing, 3rd Edition, Cambridge
  University Press, 2007.

\end{thebibliography}
	
\end{document}